\documentclass[aps,preprint,showpacs]{revtex4}
\usepackage{epsfig}

\newcommand{\ba}{\begin{eqnarray}}
\newcommand{\ea}{\end{eqnarray}}

\newcommand{\be}{\begin{equation}}
\newcommand{\ee}{\end{equation}}

\newcommand{\bra}[1]{\langle  {#1}  \vert }
\newcommand{\ket}[1]{\vert {#1} \rangle }

\begin{document}

\title{
%
%
\[ \vspace{-2cm} \]
\noindent\hfill\hbox{\rm  } \vskip 1pt
\noindent\hfill\hbox{\rm SLAC-PUB-9224} \vskip 1pt
%
On Unitary Evolution of a Massless Scalar Field In A Schwarzschild
Background: Hawking Radiation and the Information Paradox
}

\author{Kirill Melnikov and Marvin Weinstein}
\affiliation{
Stanford Linear Accelerator Center\\
Stanford University, Stanford, CA 94309\\
E-mail: melnikov@slac.stanford.edu, niv@slac.stanford.edu}

\begin{abstract}
We develop a Hamiltonian formalism which can be used to discuss the
physics of a massless scalar field in a gravitational background of
a Schwarzschild black hole.  Using this formalism we show that the time
evolution of the system is unitary and yet all known results such as
the existence of Hawking radiation can be readily understood.
We then point out that the Hamiltonian formalism leads to interesting
observations about black hole entropy and the information paradox.
\end{abstract}


\maketitle

\section{Introduction}

Hawking's 1974 paper \cite{hawking} triggered great interest in
both the existence of the radiation which bears his name and
speculations as to what his result implied for the validity of
quantum field theory in a black hole background.  Subsequent
calculations, which showed that the phenomenon was robust
\cite{unruh,jacobson,parikh} and supported the view that the
radiation appeared to be completely thermal, only added to this
interest.  Combining these results with speculations as to what
happens when the black hole evaporates led Hawking and others
to argue that the behavior of this system must be inconsistent
with the unitary time evolution of the underlying field theory,
since it starts out in a pure state and evolves into a thermal
ensemble at a temperature $T=1/(8\pi GM)$.

A related issue, the so-called {\it information paradox\/},
arose after Bekenstein \cite{bekenstein} argued that a black hole
of mass $M$ has an entropy proportional to its area. Here, the
main question is ``What happens to information which has already
crossed the horizon when the black hole evaporates?''.

Both questions led to suggestions that something goes
wrong with quantum field theory, even for a large semi-classical
black holes, and that a careful study of questions related to
these aspects of black hole physics would point the way to the
theory which must replace it.  Intrigued by this idea and convinced
that at least some of these questions would be easier to analyze
if we could canonically quantize the field theory in the
background of a large Schwarzschild black hole, we decided to do
just that.

In this paper we present a comprehensive treatment of the work
first discussed in an earlier Letter \cite{bhlett}, wherein we
canonically quantized  a massless scalar field in the background
of a large Schwarzschild black hole and derived the familiar
Hawking results.  In addition to a more detailed treatment of the
problem we show that our approach leads to a different picture of
Bekenstein's derivation of the entropy of a black hole and a
surprising, but self-consistent, resolution of the information
paradox.  These results suggest that despite expectations,
studying the problem of Hawking radiation at the purely semi-classical
level is unlikely, in itself, to produce new insights into the
question of how a quantum theory of gravity should behave.

\section{Preliminaries}

Before going further it is useful to review why a
Hamiltonian formulation of the problem of a massless scalar field
in the presence of a black hole background appears problematic.
Let us begin by considering the problem of
a massless scalar field with Lagrange density
\begin{equation}
    {\cal L} = \sqrt{-g} \left[ g^{\mu \nu} \,\partial_\mu \phi(x)\,
     \partial_\nu \phi(x) \right]
\label{masslesslag}
\end{equation}
in the background of a Schwarzschild black hole of mass $M$.  In
the usual Schwarzschild coordinates the metric $g_{\mu \nu}$ takes
the familiar form
\begin{equation}
    ds^2 =
- (1-{2M\over r} )\,dt^2
+  (1- {2M\over r} )^{-1}\,dr^2 + r^2 d\Omega^2 ,
\label{schwarz}
\end{equation}
where we have set Newton's constant, $G$, to one.

As is well known, the apparent metric singularity at $r=2M$ is a
coordinate artifact and, as such, does not pose a problem. The
true issue for canonical quantization is that we need to define a
family of spacelike slices which foliate the spacetime in order to
define initial data and form the Hamiltonian. Inspection of
Eq.(\ref{schwarz}) shows that surfaces of constant Schwarzschild
time change from spacelike to timelike at the horizon ($r=2M$) and
so they do not fulfill our requirements.  As we show in the next
Section, changing to Painlev\'e coordinates both eliminates the
coordinate singularity at $r=2M$ and allows us to analytically
define a satisfactory family of spacelike slices. Since, however,
the form of the metric in Painlev\'e coordinates is not well
suited to simple canonical quantization, we need to introduce yet
another transformation, to Lema\^itre coordinates, to facilitate
the quantization procedure. The Hamiltonian constructed in this
way explicitly depends upon Lema\^itre time, which is
a specific manifestation of the general theorem that ``the
Schwarzschild metric does not admit a global timelike Killing
vector field''.  Fortunately, having an explicitly time dependent
Hamiltonian is no barrier to unitary time evolution; in fact, this
is always the case for the interaction representation. The crucial
requirement is not that the Hamiltonian is time independent, but
rather that there exists a one-parameter family of unitary
operators $U(\lambda)$ which satisfy the equation
\begin{equation}
    {d U(\lambda) \over d\lambda } = -i\,H(\lambda)\,U(\lambda) .
\end{equation}

The lesson we learn from the fact that the Hamiltonian explicitly
depends upon time is that we shouldn't be looking for static
quantities such as {\it the vacuum state of the theory\/} but
rather for steady state phenomena such as the Hawking radiation.
In a sense, once we have observed that the Hamiltonian is time
dependent there is no longer a puzzle as to why Hawking radiation
can exist.  What remains is to fill in the details and show how
explicit calculations in this canonical framework lead to
Hawking's results.  We do this in the next few sections. Once the
equivalence of our discussion to previous approaches is
established we turn to a discussion of what our approach has to
say about the question of black hole entropy and the information
paradox.

\section{Coordinate Systems}

\begin{figure}
\begin{center}
\leavevmode
\epsfig{file=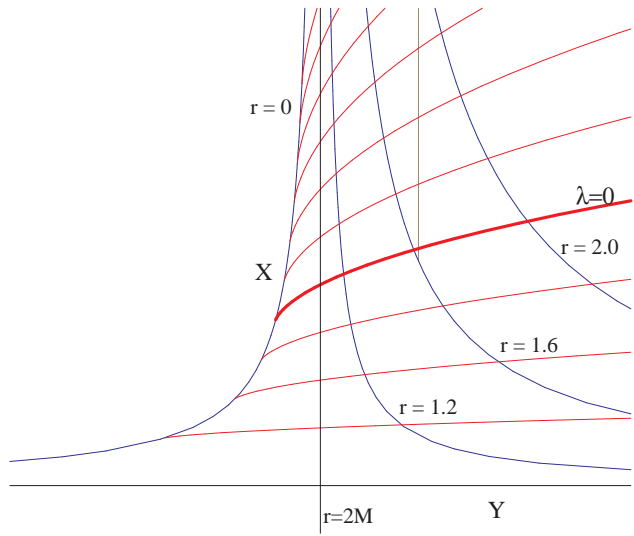,width=3.25in}
\end{center}
\caption[kruskaletc]{Only two of usual four X-Y Kruskal quadrants
are plotted and the figure
is rotated by $45^{\circ}$ to emphasize
we are studying the region from $r=0$ to $r=\infty$.  Vertical
and horizontal lines are null-geodesics.
The nearly, but not quite, horizontal curves are globally spacelike
surfaces of constant Painlev\'e time.  The hyperbolas are surfaces of
constant Schwarzschild $r$.  Lines of constant Schwarzschild time would be
straight lines originating at $X=Y=0$.}
\label{kruskal}
\end{figure}

Although we are ultimately interested in the surfaces defined by
constant Painlev\'e time, it is convenient to begin by introducing
Kruskal coordinates.  First, because these coordinates make it
particularly easy to draw null-geodesics (they are simply lines
parallel to either the $X$ or $Y$ axes shown in
Fig.\ref{kruskal});  second, because they allow us to easily
compare surfaces of fixed Schwarzschild time to surfaces of fixed
Painlev\'e time.

It is convenient to introduce dimensionless versions of $r$ and
$t$ by rescaling $ r \rightarrow 2Mr $ and $ t \rightarrow 2M t $.
Using these variables we introduce the Kruskal coordinates $X$ and
$Y$ by the equations:
\begin{equation}
 X Y = (r-1)\,e^{r}, \qquad  \qquad
        {X \over \vert Y \vert} = e^{t_S} .
\label{kruskaleq}
\end{equation}
In these coordinates the Schwarzschild metric takes the form
\begin{equation}
 ds^2 = {32\,  e^{\vert -r \vert} dX\, dY \over r} + r^2 d\Omega^2 .
\end{equation}
Eq.(\ref{kruskaleq}) tells us that fixed
Schwarzschild $r$ is a hyperbola in the $X,Y$-plane, as shown in
Fig.\ref{kruskal}, and that a surface of fixed Schwarzschild time corresponds
to a straight line $X=\vert Y \vert\,e^{t_S}$ (such lines are not
shown in Fig.\ref{kruskal}).

Painlev\'e coordinates are derived from Schwarzschild coordinates
by making an $r$-dependent shift in Schwarzschild time; i.e.,
\begin{equation}
 t = \lambda - 2\sqrt{r} - \ln\left(\left\vert{\sqrt{r}
- 1 \over \sqrt{r} + 1}\right\vert\right).
\end{equation}
This equation makes it easy to compute surfaces of fixed
Painlev\'e time.  These surfaces are the almost horizontal curves shown in
Fig.\ref{kruskal} and they clearly foliate the spacetime.  Note that while
Schwarzschild $t$ and Painlev\'e $\lambda$ differ by a function of
$r$, two events having the same $r$ are separated by equal intervals
of Schwarzschild or Painlev\'e time.  In Painlev\'e
coordinates the Schwarzschild metric takes the form
\begin{equation}
ds^2 = - \left (1 - {1 \over r} \right ) \,d\lambda^2
+ {2\,d\lambda\,dr \over \sqrt{r}} + dr^2 +  r^2 d\Omega^2 .
\label{painlevemetric}
\end{equation}
\epsfverbosetrue

Although Painlev\'e coordinates are useful for defining a family of
spacelike surfaces they are not well suited for canonical quantization
because of the cross term $d\lambda\,dr$  in the metric
and the fact that lines of constant $r$ (we assume the angular
variables $\theta$ and $\phi$ are held fixed) are not everywhere timelike.
A better coordinate system can be  obtained by considering a set of
curves $(\lambda,r(\lambda))$ whose tangent vectors,
$(1, dr(\lambda)/d\lambda)$, are orthogonal to the surfaces of constant
Painlev\'e time.  Substituting this requirement into
Eq.(\ref{painlevemetric}) for the metric in Painlev\'e coordinates
we arrive at an equation which, upon integration, gives:
\begin{equation}
    r(\lambda,r_{sch}) = \left ( r_{sch}^{3/2}
-{3\over 2}\,\lambda \right )^{2/3},
\end{equation}
where $r_{sch}$ is the value of Schwarzschild $r$ at which each
curve passes through the surface defined by
$\lambda=0$.

\epsfverbosetrue
\begin{figure}
\begin{center}
\leavevmode
\epsfig{file=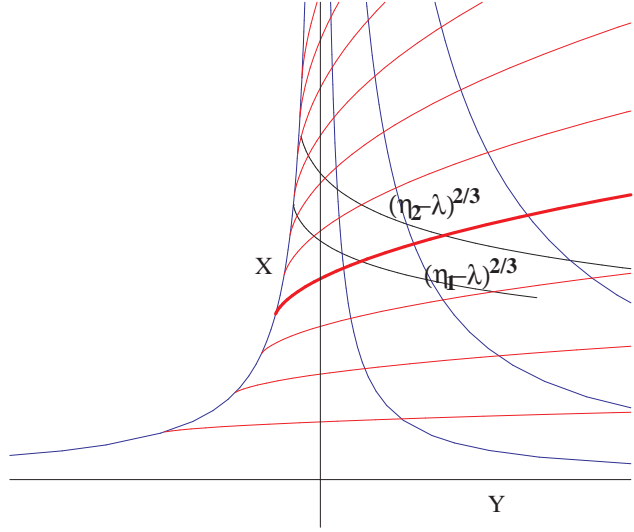,width=3.25in}
\end{center}
\caption[lemaitretime]{This is the same plot of surfaces of constant Painlev\'e time
overlaid with curves showing $r(\lambda,\eta)$ for two different initial values
of $\eta$.  Note that all such lines finally intersect the spacelike curve $r=0$}
\label{lemaitretime}
\end{figure}

We could use this coordinate system but to avoid dealing with
factors of $r^{3/2}$ it is convenient to make one more change of
variables.  This leads to Lema\^itre coordinates, which are related
to Painlev\'e $\lambda$ and $r$ by
\begin{equation}
    r(\lambda,r_{sch}) = \left (
r_{sch}^{3/2} - {3\over 2}\lambda \right )^{2/3}
    = \left ( {3\over 2} (\eta - \lambda) \right )^{2/3}.
\end{equation}
In Lema\^itre coordinates the metric takes the form
\begin{equation}
ds^2 = - d\lambda^2 + {1\over r(\lambda,\eta)} d\eta^2
+ r(\lambda,\eta)^2 d\Omega^2.
\end{equation}
It is manifestly free of coordinate singularities at $r=1$, has
no cross terms in $d\lambda$ and $dr$ and allows a completely
straightforward canonical quantization procedure.
Fig.2 shows lines of constant $\eta$ overlayed on the surfaces of
constant Painlev\'e time.

\section{Canonical Quantization}

The Lagrangian for a massless scalar field theory has the general
form given in Eq.(\ref{masslesslag}). Since the metric in
Schwarzschild, Painlev\'e and Lema\^itre coordinates is
rotationally invariant we can study the scalar field theory for
each angular momentum mode separately.  Furthermore, we are free
to expand the field $\phi(\lambda, \eta, \theta, \phi)$ in
spherical harmonics in $\theta$ and $\phi$ and restrict attention
to the $L=0$ mode, since this exhibits all of the interesting
behavior. The Lema\^itre coordinate form of the $L=0$
scalar field Lagrangian is
\begin{equation}
{\cal L} = \sqrt{-g} \,{1\over 2} \left[ (\partial_\lambda \phi_0\,(\lambda,\eta))^2
- r\,(\partial_\eta\phi_0\,(\lambda,\eta))^2 \right]
\end{equation}
where the determinant $\sqrt{-g}$ is
\begin{equation}
\sqrt{-g} = r^{3/2} = {3\over 2} (\eta-\lambda).
\end{equation}

Following the usual rules for canonically quantizing such a theory
we see that the momentum conjugate to the field is
\begin{equation}
\pi_0(\lambda,\eta) =  {3\,(\eta - \lambda)\over 2}\,\partial_{\lambda}\phi_0(\lambda,\eta ) ,
\end{equation}
and the canonical Hamiltonian is
\begin{equation}
H(\lambda) = {1\over 2} \int_{\lambda}^\infty d\eta\,
\left( {2\,\pi_0(\lambda,\eta)^2 \over 3(\eta - \lambda)} + {3\over 2} r\,(\eta-\lambda)
(\partial_\eta\phi_0(\lambda,\eta))^2 \right) .
\label{hamone}
\end{equation}
The commutation relations for $\phi_0$ and $\pi_0$ are
\begin{equation}
\left[ \pi_0(\lambda,\eta), \phi_0(\lambda,\eta') \right] = -i\,\delta(\eta - \eta').
\end{equation}

As we already pointed out, there is a one parameter family
of unitary operators, $U(\lambda)$, which satisfy the equation
\begin{equation}
     {d \over d\lambda} U(\lambda) = -i\,H(\lambda)\,U(\lambda),
\end{equation}
whose solution is the path ordered exponential
\begin{equation}
    U(\lambda) = {\cal P}\left(e^{-i\,\int_0^{\lambda}\,d\tau H(\tau) } \right) .
\end{equation}
Given these operators we define fields at later times as
\begin{eqnarray}
\phi_0(\lambda,\eta) = U(\lambda)\, \phi_0(\eta) \,U^{\dagger}(\lambda),~~~
\pi_0(\lambda,\eta)  = U(\lambda) \, \pi_0(\eta) \, U^{\dagger}(\lambda) .
\end{eqnarray}
It follows from the canonical commutation relations that
these operators satisfy Heisenberg equations of motion of the form
\begin{equation}
\label{scheulerlag}
\partial_{\lambda} \left[ (\eta - \lambda) \partial_{\lambda}\phi_0 \right]
-\partial_{\eta}\left[ (\eta - \lambda)\,r\,\partial_{\eta}\phi_0 \right] = 0.
\end{equation}

Clearly we have two options open to us. The first is to
diagonalize the Hamiltonian $H(0)$ and then explicitly construct
$U(\lambda)$.  This is at best cumbersome. The second, more
tractable option for a system with a time-dependent Hamiltonian,
is to solve the Heisenberg equations of motion and compute all
physical quantities by evaluating expectation values of the
interesting time dependent operators in a fixed initial state. We
will adopt the second approach and use the vacuum state of
the Hamiltonian $H(0)$ as our initial state.  It will be apparent
from the computations which follow that except for transient
effects, it would make no difference if we chose as our initial
state any state whose energy differed from the $H(0)$ vacuum state
energy by any finite amount. Note that there is a subtlety associated
with the computation of the time-dependent Hamiltonian at $\eta =
\lambda$ or $r(\lambda,\eta)=0$.  We will return to this point after
our discussion of the geometric optics solution of the Heisenberg
equations and the derivation of Hawking radiation.  This issue is
important and relates to the way in which the theory deals with
the {\it information paradox\/}.

Before discussing the solution of the Heisenberg equations
of motion let us  point out that it is simple to find
all of the eigenstates of $H(0)$ because it is just
a free field Hamiltonian in disguise. To see this we only need to
change variables, back to Schwarzschild $r$, using $\eta=(2/3) r^{3/2}$
and then rescale the fields by
\begin{eqnarray}
    \pi_0(r) = \sqrt{r}\,\pi_1(r),~~~
\phi_0(r) = {\phi_1(r)\over r} .
\end{eqnarray}
This converts Eq.(\ref{hamone}) to
\begin{equation}
H(0) = {1\over 2} \int_{0}^\infty dr\,
\left( \pi_1(r)^2  + r^2(\partial_r\,{\phi_1\over r})^2
\right) ,
\label{hamtwo}
\end{equation}
which  is the Hamiltonian of the $L=0$ mode of a free
massless field in flat space.  One constructs the eigenstates of
this Hamiltonian in the usual way by expanding the fields in terms
of annihilation and creation operators:
\begin{eqnarray}
    \phi_1(r) =
\int_0^{\infty} {d \omega \over \sqrt{\pi\,\omega }}\,\sin(\omega r)
    \left( a^{\dag}_\omega + a_\omega \right),~~~
 \pi_1(r) =
i \int_0^{\infty} d\omega \,\sqrt{ \omega \over \pi}\,\sin(\omega r)
    \left( a^{\dag}_\omega - a_\omega \right) .
\label{ffexpansion}
\end{eqnarray}
and defining the vacuum state $\ket{0}$ to be the state that is
annihilated by all of the $a_\omega$'s.

\section{A Look Ahead}

At this point we have to deal with two important issues. The first
is to check how the formalism just presented leads to the usual
Hawking results.  The second is to see how this Hamiltonian
approach modifies the way we think about the questions of black
hole entropy and the information paradox.

Since one of the main arguments for the usual interpretation of
Hawking radiation is that a black hole is a thermodynamic system,
before discussing the case of the black hole it is worth spending
some time reviewing the simpler problem of a constantly
accelerating mirror in a $1+1$ dimensional spacetime. We include
this discussion to emphasize that a manifestly non-singular
quantum mechanics problem with a time-dependent Hamiltonian and
unitary time evolution can produce, what looks to some observers,
like purely thermal outgoing radiation.  Furthermore, although our
discussion will cover well worn ground, the technique we use to
derive this result differs somewhat from the approach presented in
Ref. \cite{birel}.  In particular, we will focus on a simple
configuration space solution of the field equations, relating all
measurements back to the initial surface on which we quantized the
theory (i.e., at time $\lambda=0$). This allows us to get a
better handle on the physical assumptions being made when we
discuss the initial state.

In Section\ref{sectionmovingmirror} we set up the problem of
the moving mirror and discuss the solution of the field equations
in the same way we will discuss them for a field theory in a black
hole background.  One reason for studying this problem
in detail is that later, after discussing the eternal black hole problem
as it is usually formulated, we will reformulate it in the spirit
of the moving mirror problem in order to justify our choice of the
{\it vacuum state\/} on physical grounds.

After setting up the moving mirror problem we discuss what an
Unruh thermometer would measure at late times, showing that the
process of adiabatically turning on and then turning off such a
device leads to a measurement of what appears to be a
non-vanishing temperature.  Proceeding in the same vein we give a
detailed derivation of the energy flux passing through a fixed
position in space and show that it  appears to be thermal.  We do
this to emphasize the fact that a perfectly unitary quantum system
can exhibit some apparently peculiar behavior if the Hamiltonian
is time dependent.  Technically this computation is similar to the
computation one has to do for the black hole case.  The most
important result, for both cases, is that although the energy
density of the outgoing radiation is divergent and therefore ill
defined, the flux computation is free of divergences and unique.

These arguments are followed by Section \ref{sectionbogol}, where we
attempt to match our approach to the Bogoliubov transformation
technique, often discussed in the context of black hole physics.
Here we show that while one can establish such a connection,
matching the Bogoliubov approach onto a Hamiltonian formalism that
explicitly works with finite times is somewhat unnatural.

Finally, we conclude our excursion into theories in flat space
with a variant of the moving mirror problem where there is both a
moving and a fixed mirror.  This problem is interesting because it
exhibits a peculiar feature of the late time problem; namely, that
at late times, the fields over most of space depend only upon
degrees of freedom which, on the original surface of quantization,
are localized within an exponentially small region surrounding a
single point.  This strange behavior is really just a reflection
of the causal structure of the problem. It is a feature of the
Schwarzschild problem as well and will play a role in our
discussion of Bekenstein entropy.

With our foray into non-gravitational physics behind us, we
turn our attention to the case of the Schwarzschild black
hole.  A brief statement of what we mean by the {\it geometric
optics\/} approximation to the field equations is followed by
a recapitulation of the Unruh thermometer and energy flux computations
for the black hole.  Once we have shown that our approach reproduces
the well known results, we turn to things which can be better discussed
in this framework.

The first benefit which follows from discussing the black hole
problem between surfaces separated by finite times is that we can
discuss a variant of the problem of Hawking radiation which
provides a rationale for picking a particular vacuum state. For
this purpose we consider the problem of an infalling mirror,
i.e., a version of the black hole problem in which, up to a time
$\lambda_0$ the black hole is surrounded by a reflecting sphere of
large, fixed radius $R_0$, which at time $\lambda_0$ falls into
the black hole along a Lema\^itre timeline.  By a reflecting
sphere we mean that we assume that the field always vanishes on
and inside the surface.  Since the field is in a region of
vanishingly small gravitational field for an infinite time in the
past, at least if $R_0$ is chosen large enough, we can argue that
from the point of view of a local observer in this region it is
sensible to assume that the system starts out  in the vacuum
state. We then show how, in a manner very reminiscent of
calculations done for a self-assembling black hole, the Hawking
radiation forms as the mirror approaches the horizon.

Another problem which becomes more approachable because we work
between finite times is the so-called back reaction problem. The
back reaction problem is equivalent to the observation that the
eternal Schwarzschild background is not consistent with the
addition of the scalar field theory, since the energy momentum
tensor we compute for the scalar field has a uniquely defined
non-vanishing flux term and the Einstein tensor for the
Schwarzschild solution vanishes.  Since we can discuss this issue
for large black holes and finite times, during which the mass of
the hole does not change much, we argue it is possible to ask and
solve the question of what a truly self-consistent problem would
look like.  We emphasize that this is possible because we avoid
the issue of what happens at infinite times in the future when the
evaporation process becomes rapid and runs to completion.

To discuss the question of black hole entropy we study a variant
of the black hole problem in which we place a large reflecting
surface around the black hole, but now we assume that the field
theory exists only inside this surface.  We then show that to an
outside observer a body constructed in this way appears to be a
thermal system with the familiar Bekenstein entropy, however when
we look inside we see it is not an equilibrium system.  We discuss
the meaning of this observation.

Finally, we turn to the so-called information paradox. We argue
that the fact that the geometric optics solution is exact for the
two-dimensional black hole tells us that we really have to take
into account the spacelike line, $r=0$, stretching from the quantization
surface $\lambda_0$ to the surface on which we do measurements. To
do this in a consistent way for two and four dimensions we
introduce a lattice in Lema\^itre coordinate $\eta$ and
demonstrate that the spectrum of the time-dependent Hamiltonian is
constantly changing; this proves that the time dependence of the
problem is not a coordinate artifact.  We conclude with a
discussion of the picture which is suggested by this analysis;
namely, that a very weakly coupled remnant forms as the black hole
evaporates.

\section{The Moving Mirror}
\label{sectionmovingmirror}

Consider a free field theory in flat space, together with the
boundary condition that the field vanishes on and to the left of a
curve $x(t)$.  To discuss the solution of the Heisenberg field
equations for this problem it is helpful to review
the simplest way of solving the free field Heisenberg equations
when there are no boundary conditions.  Start from the free
field Euler-Lagrange equation
\begin{eqnarray}
(\partial_t^2 - \partial_x^2 )\,\phi(t,x) = 0,
\end{eqnarray}
and rewrite it as
\begin{eqnarray}
(\partial_t - \partial_x)\,(\partial_t + \partial_x)\,\phi(t,x) = 0.
\end{eqnarray}
Then observe that the general solution to this equation can be
written as:
\begin{equation}
\phi(t,x) = f(x-t) + g(x+t).
\label{genflatsol}
\end{equation}
The functions $f$ and $g$ are determined by the values of
$\phi(t,x)$ and its time derivative at $t=0$:
\begin{eqnarray}
\partial_x f(x) = {1\over 2} (\partial_x\phi_0(x) - \pi_0(x)),~~~
\partial_x g(x) = {1\over 2} (\partial_x\phi_0(x) + \pi_0(x)),
\end{eqnarray}
where $\phi_0$ and $\pi_0$ are initial conditions at $t=0$ surface.
It is a simple matter to integrate these equations to determine $f(x)$ and $g(x)$.

\epsfverbosetrue
\begin{figure}
\begin{center}
\leavevmode
\epsfig{file=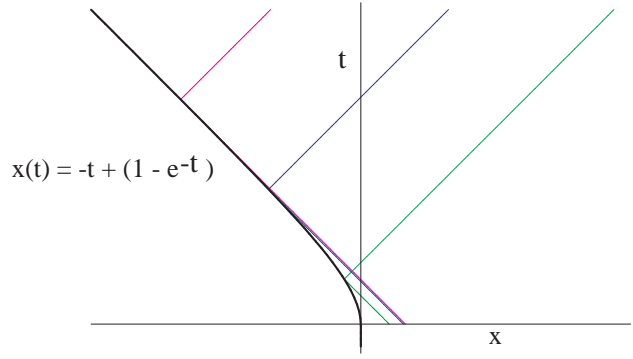,width=3.25in}
\end{center}
\caption[movingmirror]{A plot of the moving mirror and three null
geodesics which represent contributions to $g_0(x-t)$.  These curves
are selected to show that null lines which hit the mirror at later
times come from points which start out closer and closer to the point
$x=1$.}
\label{movingmirror}
\end{figure}

Next consider, as shown in Fig.\ref{movingmirror}, the case of a
field theory with moving boundary $x(t) = -t + A\,(1-e^{-2\,\kappa
t})$, where we have chosen to plot the curve for $A=1$ and
$\kappa=1/2$.  Clearly this system has a time-dependent
Hamiltonian and, nevertheless, possesses a unitary time
development operator.  If, as shown, we assume that before $t=0$
the mirror is at rest and has been that way for an infinite amount
of time, then it is reasonable to assume that the initial state of
the problem at $t=0$ is the vacuum state for the free field theory
defined by the condition $\phi_0(0)=0$.

In this case the Euler-Lagrange equations remain unchanged,
however the solution needs to be modified to maintain the boundary
condition which says that $\phi(t,x(t)) = 0$. This is easily done
by adding a reflected wave $g_0(x-t)$ to the general solution
, so that it becomes:
\begin{eqnarray}
\phi(t,x) = \theta(x-t)\,f(x-t) + g(t+x) + \theta(t-x)\,g_0(x-t).
\label{phievalone}
\end{eqnarray}
As in the case with no boundary conditions we determine $f(x)$ and
$g(x)$ from the initial data on the $t=0$ surface:
\begin{eqnarray}
 g(x) &=&
{1 \over 2}\,\int_0^x d\zeta \left( {d\phi(\zeta) \over d\zeta} + \pi(\zeta)
 \right) \label{gofx}, \\ \nonumber
 f(x) &=& {1 \over 2}\,\int_0^x d\zeta
\left( {d\phi(\zeta) \over d\zeta} - \pi(\zeta) \right).
\label{fofx}
\end{eqnarray}
Given this, we find from the requirement that the total
field $\phi(t,x)$ vanishes on the curve $x(t)$ that $g_0$
is given by
\begin{equation}
  g_0(x-t) = - g(x_0(t,x)),
\label{phievaltwo}
\end{equation}
where, as indicated in Fig.\ref{movingmirror}, $x_0(t,x)$ is the
point on the $t=0$ surface from which the reflected wave came.
The function $x_0(t,x)$ is determined by observing that the light
ray which comes to the point $(t,x)$ left the mirror at some point
$(t_1,x(t_1))$ and that the incident wave which arrived at this
point came from $x_0(t,x)$.  These statements are equivalent to
the equations
\begin{eqnarray}
  x - t = x(t_1) - t_1,~~~
  x(t_1)+t_1 &=& x_0(t,x),~~~
  x(t_1) = -t_1 + A(1 - e^{-2\kappa t_1}). \label{ethree}
\end{eqnarray}
From  Eq.(\ref{ethree}) we immediately obtain
\begin{equation}
    x_0(t,x) = A(1 - e^{-2\kappa t_1})
\label{xzero}
\end{equation}
or, equivalently,
\begin{equation}
       -2 t_1 = {1 \over \kappa}\,\log \left( {A - x_0(t,x) \over A} \right).
\end{equation}
Combining this with Eq.(\ref{ethree}) we derive
\begin{eqnarray}
\label{xnoughttwo}
  x - t &=& - 2 t_1 + A (1 - e^{-2 \kappa t_1})
= - 2 t_1 + x_0(t,x), \nonumber \\
  x-t &=& x_0(t,x)
+ {1 \over \kappa } \log \left( {A - x_0(t,x) \over A } \right),
\end{eqnarray}
which, for $t-x \gg A$, has the approximate solution,
\begin{equation}
  x_0(t,x) \simeq A ( 1 - e^{-\kappa (t-x) } ).
\label{xnought}
\end{equation}

This completes our solution of the Heisenberg equations of motion
for the free field in the presence of a moving mirror.  In the
sections to follow we will see that because the point $x_0(t,x)$
becomes arbitrarily close to the point $A$ for $ t \gg x$, a
thermometer placed at a distance from the mirror will measure a
temperature $ \kappa/2 \pi$ and a detector will see an outgoing
energy flux $\kappa^2 / 48 \pi$.

\section{Unruh thermometer}
\label{sectionunruh}

We begin with a precise definition of what we mean by a
thermometer. In what follows we will take {\it a thermometer\/} to
be a simple quantum system, with multiple energy levels, locally
interacting with the field $\phi(t,x)$.  Two terms should be added
to the massless scalar field Lagrangian in order to specify the
interaction of the thermometer with the field.  The first is a term
which defines the eigenstates of the non-interacting thermometer
and the second, is an interaction term of the form
\begin{equation}
V_{\rm int}(t) = \epsilon\,e^{-(t-t_0)^2/2\sigma} Q\,\phi(x,t).
\end{equation}
Here $\epsilon$ is the small parameter in which we will perturb,
$t_0$ and $\sigma$ define the range in $t$ for which the
interaction is turned on and $x$ specifies the spatial location of
the thermometer. The operator $Q$ is assumed to be an operator
which causes transitions among the energy eigenstates of the
thermometer.

A number of  assumptions have to be made in order to get
reasonable results. First, in order for the thermometer to know
the mirror is moving, it is necessary to assume that $t_0 \gg x$.
Second, we must impose an adiabatic condition, $\sqrt{\sigma} \gg
1/E$, where $E$ is the typical excitation energy of the
thermometer, so that we do not excite the thermometer just by
turning it on or off.  Finally, we must impose the condition $ E
\sim \kappa $, so that the acceleration of the mirror is capable
of exciting the higher states of the thermometer.

Given these assumptions, second order perturbation theory in
$\epsilon$ tells us that the probability of the thermometer being
excited to a state with energy $E$ is
\begin{eqnarray}
{\cal P}(E,E_0) &=& \epsilon^2 \vert \bra{E} Q \ket{E_0} \vert^2
\int dt\,dt'
e^{-i(E-E_0)(t-t')-[(t-t_0)^2+(t'-t_0)^2]/2\sigma}
 \langle \phi(t,x) \phi(t',x)\rangle,
\label{pertmirror}
\end{eqnarray}
where $\langle {\cal O}\rangle$ stays for the vacuum expectation value
of the operator ${\cal O}$.

We must now evaluate $\langle \phi(t,x) \,
\phi(t',x)\rangle$ using Eqs.(\ref{phievalone}-\ref{phievaltwo}),
which tell us how to rewrite $\phi(t,x)$ and $\phi(t',x)$ in terms
of $\phi(x)$ and $\pi(x)$ on the surface $t=0$.  Once we have done
this we can rewrite these $t=0$ operators in terms of
annihilation and creation operators and evaluate the resulting
expression. We obtain
\begin{eqnarray}
{\cal P}(E,E_0) =  {\epsilon^2 \sqrt{\pi} \over 2}\,
{\vert \bra{E} Q \ket{E_0} \vert^2 \over E-E_0}
\times \left[ {1 \over e^{2\pi\,(E-E_0)/\kappa} -1} \,\right] .
\end{eqnarray}
This result shows that the thermometer reacts as if it is in
interaction with a heat bath at a temperature $\kappa/2\pi$. From
this point on we will assume, without loss of generality, that
$E_0=0$ in order to simplify the equations.

Let us now discuss the details of the calculation.
To obtain Eq.(\ref{pertmirror}) we must first compute the quantity
\begin{equation}
G(t,t') = \langle \phi(t,x) \, \phi(t',x)\rangle,
\end{equation}
when both $t$ and $t'$ are much greater than $x$.  When these
inequalities are satisfied we see from Eq.(\ref{phievalone}) that
$\phi(t,x)$ is given by
\begin{equation}
\phi(t,x) = g(t+x) - g(x_0(t,x)).
\label{phiI}
\end{equation}

It follows from Eq.(\ref{gofx}) that
\begin{equation}
\phi(t,x) = {1 \over 2}\,\int_{x_0(t,x)}^{t+x} d\zeta \left( {d\phi(\zeta) \over d\zeta}
    + \pi(\zeta) \right).
\end{equation}
Using the expansion of the operators $\phi(\zeta)$ and
$\pi(\zeta)$ defined
in Eq.(\ref{ffexpansion}) we obtain
\begin{equation}
{d\phi(\zeta) \over d\zeta} + \pi(\zeta) = \int_0^{\infty} d\omega\, \sqrt{\omega \over \pi}\,
\left( e^{i\omega\zeta}\,a^{\dag}_{\omega} + e^{-i\omega\zeta}\, a_{\omega} \right)
\label{ffexpansiontwo}
\end{equation}
from which it follows that
\begin{eqnarray}
\phi(t,x) = {i \over 2} \int_0^{\infty} d\omega \,{1\over \sqrt{\pi \omega}}
\,\left[ \left(e^{i\omega x_0(t,x)}-e^{i\omega (t+x)} \right)\,a^{\dag}_{\omega}
-\left(e^{-i\omega x_0(t,x)}-e^{-i\omega (t+x)}\right)\,a_{\omega} \right].
\label{phioftandx}
\end{eqnarray}
It is straightforward to evaluate the expectation value
\begin{eqnarray}
\langle \phi(t,x)\phi(t',x) \rangle
=-{1 \over 4\pi} \int \limits_{0}^{\infty} {d\omega \over \omega}
 \left( e^{i\omega(x_0(t'x)-x_0(t,x))}
 + e^{i\omega(t'-t)}
- e^{i\omega(x_0(t',x) -t-x)} - e^{i\omega(t'+x-x_0(t,x))} \right)
\nonumber\\
\label{fourterms}
\end{eqnarray}
Computing these integrals we arrive at the result
\begin{eqnarray}
\langle \phi(t,x) \phi(t',x) \rangle &=& {-1\over 4\pi}
\left[ \ln(x_0(t',x) -x_0(t,x) + i\eta)
 + \ln(t'-t + i\eta)
\right. \nonumber\\ && \left.
 - \ln(x_0(t',x)-t-x + i\eta)
 - \ln(t'+x -  x_0(t,x) + i\eta) \right],
\label{fourlogs}
\end{eqnarray}
where an infinitesimal positive quantity $\eta$ has
been introduced.
Substituting this into Eq.(\ref{pertmirror}), we see that the
last two terms in Eq.(\ref{fourlogs}) are  damped
by a factor of $e^{-\sigma E^2}$, which is negligible thanks to
the adiabatic assumption. The reason this happens is that both $t$
and $t'$ are restricted to be near $t_0$ which is assumed to
be large enough so that $x_0(t_0,x) \simeq A$. Therefore,
either  $t$ or $t'$ integration can be done, yielding the suppression factor.
This leaves only the integration of the first two terms.
Since the first term takes the most work let us begin with
it.

The integral we must evaluate to obtain the first term's contribution
to the transition probability is
\begin{eqnarray}
{-1\over 4\pi}
\int_{-\infty}^{\infty} dt dt'\, e^{-iE(t-t')} e^{-(t-t_0)^2/2\sigma}
e^{-(t'-t_0)^2/2\sigma}  \ln(x_0(t',x)-x_0(t,x)+i\eta).
\end{eqnarray}
In order to simplify the Gaussian terms it will be convenient to
let $t \rightarrow t+t_0$ and $t'\rightarrow t'+t_0$ and then,
replacing $x_0$ by Eq.(\ref{xzero}) we obtain
\begin{eqnarray}
{-1\over 4\pi}
\int_{-\infty}^{\infty} dt dt' \,e^{-iE(t-t')} e^{-t^2/2\sigma}
e^{-t'^2/2\sigma}
\left(\ln(A) -\kappa(t_0-x) + \ln(e^{-\kappa t} - e^{-\kappa t'}) \right).
\end{eqnarray}
As in the previous discussion, we see that the $\ln(A)$ term and
the term $-\kappa(t_0-x)$ are both suppressed by $e^{-\sigma
E^2}$.  Having removed all of the $t_0$ dependence it is
convenient to define $u=t'-t$ and $v=t'+t$ and rewrite what is
left as
\begin{eqnarray}
{-1\over 8\pi} \int_{-\infty}^{\infty} dv du \,e^{iEu} e^{-u^2/4\sigma}
e^{-v^2/4\sigma}
 \left({-\kappa \over 2}(v + u) + \ln(e^{\kappa u} - 1) \right).
\end{eqnarray}
Once again all  the terms linear in $v$ and $u$ vanish
or give a contribution of order $e^{-\sigma E^2}$, so the only integral
we have to compute reads:
\begin{equation}
{-1\over 4\pi} \sqrt{\pi\sigma} \int_{-\infty}^{\infty}du \,e^{iEu} e^{-u^2/4\sigma}
\ln(e^{\kappa u} - 1).
\label{toughint}
\end{equation}

In order to handle this term we rewrite the integral in
Eq.(\ref{toughint}) as a sum of integrals over positive $u$, to
obtain
\begin{eqnarray}
{-1\over 4\pi} \sqrt{\pi\sigma}
\int_{0}^{\infty}{\rm d}u \,e^{-u^2/4\sigma}
\left[\,e^{iEu} \left(\kappa\,u + \ln(1-e^{-\kappa u}) \right)
+ e^{-iEu} \,\ln(1-e^{-\kappa u}) + i\pi \,\right].
\end{eqnarray}
At this juncture we see that if we rewrite the integral in terms
of the variable $\xi = \kappa u$ and expand the logarithmic terms
in a series in $e^{-\xi}$, we obtain
\begin{eqnarray}
{-1\over 4\pi} {\sqrt{\pi\sigma}\over \kappa}
\int \limits_{0}^{\infty}{\rm d}u  \,e^{-\xi^2/4\sigma\kappa^2}
\left[ e^{i E\xi/\kappa} \xi - \sum_{n=1}^{\infty}\,
{e^{(iE/\kappa - n)\xi} \over n}
 - \sum_{n=1}^{\infty} {e^{-(iE/\kappa+n)\xi}\over n} \right].
\end{eqnarray}
The condition $\sigma\kappa^2 \gg 1$ allows us to replace the term
$e^{-\xi^2/4\sigma\kappa^2}$ by unity, which then allows us to
carry out all of the integrations and obtain
\begin{equation}
{\sqrt{\pi\sigma} \over 4\pi\kappa }\left[{\kappa^2 \over E^2} + \sum_{n=1}^{\infty}
{2\over n^2 + \left({E\over \kappa}\right)^2} \right] - {\sqrt{\pi\sigma}\over 4 E}.
\end{equation}
The sum can be explicitly done using the identity
\begin{equation}
\sum_{n=1}^{\infty}{1\over n^2 + \left({E\over \kappa}\right)^2} =
{\pi\kappa \over 2 E} \left[ {2 \over e^{2 \pi E /\kappa} -1}
+ 1 - {\kappa \over \pi E} \right],
\end{equation}
which makes the total undamped contribution of this term to the
transition matrix element
\begin{equation}
{ \sqrt{\pi \sigma}\over 2E} \left[ {1 \over e^{2 \pi E/\kappa} - 1} \right].
\end{equation}
Finally it only remains to show that the
contribution of the second term in Eq.(\ref{fourlogs}) vanishes.
This is easily done by rewriting $\xi = Eu$, using the fact that
$\sigma E^2 \gg 1$ and rewriting everything as a sum of integrals
over the range $u=0\dots\infty$.

\section{The Energy Flux}
\label{sectionenergyflux}

Having seen that a thermometer at a fixed location will, at large
times in the future, measure a temperature $T=\kappa/2\pi$, we
would like to delve further into this phenomenon.  One way to do
this is to compute the net flux of energy passing through the
point $x$, to see if the thermometer is heating up due to
a flux emanating from the location of the mirror.  This means we
need to compute $T_{tx}$, the energy-momentum tensor for the
massless field.

In Minkowski space the energy-momentum tensor for the massless
scalar field is defined to be
\begin{equation}
T_{\mu \nu} = {1 \over 2} \left\{ \partial_\mu \phi(t,x),
\partial_\nu \phi(t,x) \right\},
\end{equation}
where $\{A,B\}$ denotes the anti-commutator.
In general, the expectation value of any component of $T_{\mu \nu}$
is  divergent since we have to evaluate the product of
two  quantum fields at the same spacetime point. Thus
we need  a regularization procedure
which can identify possible infinities.
In what follows we adopt a point splitting method
to regularize the stress-energy
tensor and define:
\begin{equation}
T_{tx} = {1 \over 2} \left\{ \pi(t+\delta,x),
\partial_x \phi(t-\delta,x) \right\}.
\label{tmunu}
\end{equation}
It is understood that the limit $\delta \rightarrow 0$ should be taken
at the end of the calculation. It turns out that in this limit
the energy density $T_{tt}$ diverges as $1/\delta^2$
but the flux, $T^{tx}$, is finite and unique:
\begin{equation}
T^{tx} = {\kappa^2 \over 48 \pi}.
\label{fl1}
\end{equation}

To derive Eq.(\ref{fl1}), one takes  appropriate
derivatives in Eq.(\ref{phioftandx}) and
evaluates the expectation value of the commutator defined in
Eq.(\ref{tmunu})
The result is:
\begin{eqnarray}
T_{tx} ={1 \over 4\pi}\,\left[ -{1\over \delta^2}
+ {x_0^{'}(\xi)\,x_0^{'}(\xi+\delta)
\over (x_0(\xi) - x_0(\xi+\delta))^2}
+  {x_0^{'}(\xi) \over (t+\delta+x -x_0(\xi))^2 }
- {x_0^{'}(\xi+\delta) \over
(t +x -x_0(\xi+\delta))^2} \right],
\label{geq}
\label{singularterms}
\end{eqnarray}
where $x_0^{'}(\xi) = {\rm d}x_0(\xi)/{\rm d}\xi$ and $\xi = t-x$.
Each of the first two terms in Eq.(\ref{singularterms})
diverges if $\delta \rightarrow 0$ whereas the last two terms take
finite limits and cancel one another. Substituting
the explicit form of $x_0(t,x)$ from
Eq.(\ref{xnought})
into the second term of Eq.(\ref{geq}) and expanding it in powers
of $\delta$ we find that it becomes
\begin{equation}
 {1 \over \delta^2} - {\kappa^2 \over 12} + {\cal O}(\delta).
\end{equation}
From this we see that in the limit $\delta\rightarrow 0$ all
singular terms in $\delta$ cancel, yielding the  result
\begin{equation}
T^{tx} = {\kappa^2 \over 48 \pi} .
\end{equation}
There are several features of this calculation which are generic and
are therefore worth further discussion.

The first point which merits discussion is the
finiteness of our result for the flux.
The general reason for that  is
the  theorem discussed in Ref. \cite{birel} where
divergent terms which can
appear in $T_{\mu\nu}$ for a general background gravitational
field are explicitly given.
Evaluation of these terms for the case of flat Minkowski space with
a moving boundary as well as  the Schwarzschild black hole shows
that no divergent terms can arise in the computation of the flux and
therefore it must come out finite.  The preceding discussion
shows, by explicit calculation, how this works for the case of the
moving mirror.

Next, we wish to observe that Eq.(\ref{singularterms}) shows that
only those terms which are singular in the $\delta
\rightarrow 0$ limit contribute to the result for  the flux.
A peculiar feature of  those terms is that for them
the points $(t+\delta,x)$ and $(t,x)$ can be
traced back to the same point on the initial surface. Terms which
are finite in the limit $\delta\rightarrow 0$ always cancel
exactly. One way of describing this state of affairs is that the
first term in Eq.(\ref{singularterms}) represents flux coming from
the right and the second term represents flux coming from the left
(i.e., the mirror).  In the case of Minkowski space with no
mirror, these two contributions would cancel exactly, however in
the case of the moving mirror the flux reflected from the mirror
is subjected to a time dependent redshift.  It is this time
dependent redshift which produces a non-cancelling
finite addition to the flux. As we will see in a later Section, a
similar decomposition of the terms contributing to the energy
momentum tensor is possible for the case of a black hole.  In the
Schwarzschild case the flux coming from the right behaves much
like the Minkowski space contribution, but the flux coming from
the left emanates from the vicinity of the horizon. This flux also
sees a time dependent redshift due to the fact that flux which
arrives at a slightly later time originates from a point which is
slightly closer to the horizon. Just as in the moving mirror case,
this time dependent redshift is the cause of the finite,
non-cancelling contribution to the outgoing flux which we identify
as Hawking radiation. Finally, we should point out that the same
approach can be used to describe what happens if the mirror moves
along an arbitrary trajectory $x(t)$ which asymptotes to the light
cone, since all that changes in the calculation is the way in
which one deals with the function $x_0(t,x)$.

\section{The Bogoliubov Transformation}
\label{sectionbogol}

Bogoliubov transformations are a common tool used to deal with
field theory in curved space time, in particular in cases of the
moving mirror and the Schwarzschild black hole.  In this Section
we show how to understand these ideas within the Hamiltonian
formalism.

The usual context within which one discusses Bogoliubov
transformations is a situation in which there is a time $t_1$ before
which the Hamiltonian is time independent and free, and a time
$t_2$ after which it is time independent and free.  In this case
the Bogoliubov transformation is straightforward to both define
and compute.  All one has to do is Fourier expand the fields
$\phi(t_1,x)$ and $\pi(t_1,x)$ in terms of annihilation and
creation operators $a_{\omega}(t_1)$ and $a^{\dag}_{\omega}(t_1)$,
as in Eq.(\ref{ffexpansiontwo}) and do the same for the fields
$\phi(t_2,x)$ and $\pi(t_2,x)$.  Then, if we have explicit
formulas relating the fields $\phi(t_2,x)$ and $\pi(t_2,x)$ to the
fields at time $t_1$ we get an explicit relation between
$a_{\omega}^{\dag}(t_1)$ and $a_{\omega}(t_1)$ and their
counterparts at time $t_2$.  This relationship is the desired
Bogoliubov transformation.  Applying these ideas to the
case of the moving mirror would be quite simple except for the
fact that, although before time $t_1=0$ the theory is indeed time
independent and free, there is no time $t_2$ for which the same
conditions apply. There is, however, a sense in which we can treat
the system as being almost time independent.  We can then use the
corresponding Bogoliubov transformation to compute such things as
the thermometer response and outgoing energy flux.

\epsfverbosetrue
\begin{figure}
\begin{center}
\leavevmode
\epsfig{file=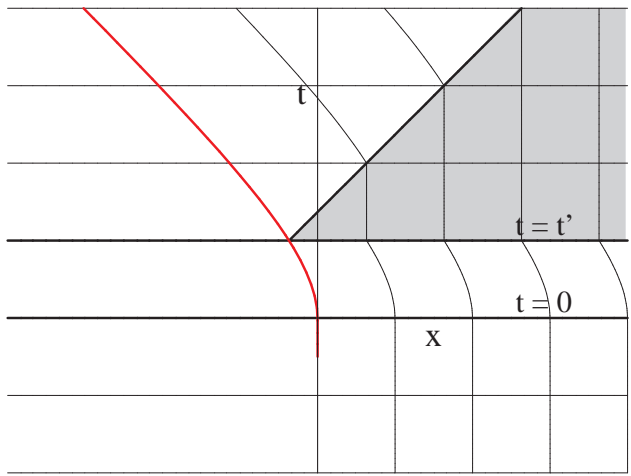,width=3.25in}
\end{center}
\caption[bogoliubov]{The coordinate system used to discuss the
Bogoliubov transformation.  Note that before time $t=0$ and
in the shaded wedge formed by the lines $t=t'$ and the $45^{\circ}$ line
we use ordinary rectangular Cartesian coordinates, but in the
remaining region we use lines which are translations of the
mirror trajectory.  For fixed $x$ the field $\phi(\tau,x)$
for all $(\tau,x)$ in the shaded wedge can be computed in terms of
its values on the line $t=t'$ using free-field equations of motion.}
\label{bogoliubov}
\end{figure}

To understand why this works consider the coordinate system shown
in Fig.\ref{bogoliubov} and focus on the point $(t',x')$.
Causality  requires that for times $t$ such that $x+t' > t > t'$
(i.e., for all points $(t,x)$ lying within the shaded triangular
wedge) the field operators $\phi(t',x')$ and
$\pi(t',x')$ evolve according to the infinite volume free field
equations.  Thus, to compute what a thermometer moving along the
timeline $(t,x')$ would see, we can expand the initial $t=0$ state
in terms of the annihilation and creation operators of the
instantaneous Hamiltonian $H(t')$ and then use infinite volume
free field equations to evaluate the relevant correlation
functions.  This approach should lead to our earlier results if,
as we will show in a moment, we work at large enough $x$. It has
(obvious) limitations, however, if we want to study this problem
at some fixed finite distance from the origin over a finite
interval in time.

From Fig.\ref{bogoliubov} we see that the farther out $x'$ is
along the $x$-axis the longer the time interval for which infinite
volume free field propagation makes sense; nevertheless, we see
that for any finite $x'$ the approximation eventually breaks down.
This means that Fourier transforming in the time variable in order
to identify the annihilation and creation operators is only an
approximation to what is going on.  One always has the option of
putting the initial surface in the infinite past and working at
$x'=\infty$, which would make the approximation exact;  but if one
is interested in a detailed picture of how the system develops in
time, working at past and future infinity is a severe limitation.
Having said that, let us show how the Bogoliubov transformation
should be understood in the Hamiltonian framework.

To carry out the Bogoliubov transformation we
begin by considering the instantaneous Hamiltonian of the system at
time $t=t'$, when the mirror is at the position $x(t')$:
\begin{equation}
H(t') = {1 \over 2} \, \int_{t'}^{\infty}\,
 dx\, \left( \pi(t',x)^2 + \partial_x\phi(t',x)^2\right).
\end{equation}
This is just the Hamiltonian of a free massless field
theory defined on the interval $x(t') \le x \le \infty$
and it is diagonalized by expanding the field and its conjugate
momentum in terms of annihilation and creation operators $b_k$ and
$b^{\dag}_k$, defined by the formulas
\begin{eqnarray}
&& \phi(t',x) = \int_{0}^{\infty} {dk \over \sqrt{k \pi}}\,
\sin(k(x-x(t'))) \, \left( b^{\dag}_k + b_k \right),\\
&&\pi(t',x) = -i\,\int_0^{\infty} dk \sqrt{k \over \pi} \,\sin(k(x-x(t')))
\,\left( b^{\dag}_k - b_k \right).
\end{eqnarray}

It will be convenient in what follows to make the coefficients
of $b^{\dag}_k$ and $b_k$ simple exponentials by taking the linear combinations
\begin{eqnarray}
\pi(t',x) + {\partial\phi(t',x) \over \partial x} &=& \int_0^{\infty} dk \sqrt{k\over \pi}
\left(b^{\dag}_k\,e^{-i k (x - x(t'))}
+  b_k\,e^{i k (x - x(t'))} \right),\\
-\pi(t',x) + {\partial\phi(t',x) \over \partial x} &=& \int_0^{\infty} dk \sqrt{k\over \pi} \left(
b^{\dag}_k\,e^{i k (x - x(t'))}  + b_k\,e^{-i k (x - x(t'))} \right).
\end{eqnarray}

The next step in computing the Bogoliubov transformation is to rewrite
$\phi(t',x)$ and $\pi(t',x)$ in terms of the $t=0$ fields.
Following our earlier discussion we see that $\phi(t,x)$ is a sum of
two terms:
\begin{equation}
\phi(t',x) = \Theta(t'-x) \,\phi_I(t',x) + \Theta(x-t')\,\phi_{II}(t',x) .
\end{equation}
and, using Eq.(\ref{phioftandx}), we obtain the following expressions for
\begin{eqnarray}
\pi_I(t',x)+{\partial\phi_I\over\partial x}(t',x) =
\int dk\,\sqrt{k\over \pi}\left[
a^{\dag}_k\,  e^{i k (x+t')} + a_k\,e^{-ik(x+t')} \right],
\end{eqnarray}
and
\begin{eqnarray}
\label{fourier}
\pi_I(t',x)-{\partial\phi_I\over \partial x}(t',x) =
{\partial x_0(t',x) \over \partial x}
\int dk\,\sqrt{k\over \pi}  \left[
a^{\dag}_k e^{i k x_0(t'x)} + a_k\,e^{-i k x_0(t'x)} \right].
\end{eqnarray}

If we consider the last of  these two equations and ignore the part of
$\phi_I(t',x)$ which does not come from the mirror, and if we assume that
$t'$ is large so that $x(t')$ is large and negative, then to a
good approximation we can extract the coefficient of $b^{\dag}_k$
by simply taking the Fourier transform
\begin{eqnarray}
\sqrt{k\over\pi} b^{\dag}_k &=& -\int {dx\over 2\pi}\,e^{-ik(x-x(t'))}
\left[\pi_I(t',x)-{\partial\phi_I\over \partial x}(t',x)\right] \nonumber\\
&=& - \int {d\omega \over 2\pi} \sqrt{\omega \over \pi} \,
\int dx\, e^{-ik (x-x(t'))}
{\partial x_0(t',x) \over \partial_x} \left(a^{\dag}_{\omega}
e^{i\omega x_0(t',x)}
+ a_{\omega} e^{-i\omega x_0(t',x)} \right).
\end{eqnarray}
Changing variables from $x$ to $x_0(t',x)$,
rewriting Eq.(\ref{xnoughttwo})
in the limit of large $t'$  as
\begin{equation}
x=t'+ {1\over\kappa}\ln({A - x_0\over A}) + x_0
\end{equation}
and letting $x_0 = A(1-\xi)$ the expression for $b^{\dag}_k$ becomes
\begin{eqnarray}
\sqrt{k \over \pi} b^{\dag}_k &=& -A\,\int {d\omega \over 2\pi} \sqrt{\omega \over \pi}\int_{0}^{1} d\xi\,
e^{-ik(t'-x(t'))}
\xi^{-ik/\kappa}
\left(a^{\dag}_{\omega} e^{iA(\omega-k)(1-\xi)} + a_{\omega}e^{-iA(\omega+k)(1-\xi)}\right).
\end{eqnarray}
Since the dominant contributions to this integral come from small $\xi$ we can extend the
$\xi$-integration to infinity and evaluate the result in terms of
Gamma functions
to obtain:
\begin{eqnarray}
\label{bogoltransform}
\sqrt{k \over \pi} b^{\dag}_k = -{A\over 2\pi}&&\kern-12pt\int d\omega\,\sqrt{\omega \over \pi}\,
e^{ik(t'-x(t'))}\,\Gamma \left(1-i{k\over \kappa} \right )\,\nonumber\\
\times \left[a^{\dag}_{\omega}\, e^{iA(\omega-k)}\right.&&\left.\kern-10pt
\left(-i\over A(\omega-k)\right)^{(1-i{k\over\kappa})}
+a_{\omega}  e^{-iA(\omega+k)}\,
\left(i \over A(\omega+k)\right)^{(1-i{k\over\kappa})} \right].
\end{eqnarray}

Given this approximate Bogoliubov transformation one usually
computes the expectation values of the number operators $\bra{0}\,
b^{\dag}_k \, b_k \ket{0}$, where $\bra{0}$ stands for the state
the system started in at time $t=0$, which we have chosen to be
the state annihilated by the operators $a_{\omega}$. Using
Eq.(\ref{bogoltransform} ), one derives
\begin{equation}
\bra{0}\, b^{\dag}_k\,b_k\, \ket{0} =
{k \over 4\pi^2\kappa}\,\left[\int {\omega d\omega
\over (\omega+k)^2}\right]\,
{1 \over (e^{2\pi k/\kappa}-1)}.
\end{equation}
Except for the divergent integral in front of the expression, this
is what we need to plug into the formula for the transition
probability for the thermometer to obtain its response. The usual
way of arguing away the divergent prefactor is that the number of
particles grows with time and that one should divide by this
logarithmically divergent term to get a number of particles per
unit time.  If one does this then one can use the results of this
approximate calculation to derive the response of a thermometer
which is switched on and off for a finite period, or the flux of
energy through a given point.

To conclude this Section we would like to point out that
we do not see any advantage, either computational or conceptual,
in discussing particular choices of the vacuum states and the associated
Bogoliubov transformations for this explicitly time dependent
problem.  The time dependent nature of the problem, which
explicitly manifests itself in the time dependence of the
Hamiltonian, naturally leads
to a long time steady state behavior of physical system that
is practically independent of the initial conditions.

\section{Two Mirrors:  A Curious Phenomenon}

\begin{figure}
\begin{center}
\leavevmode
\epsfig{file=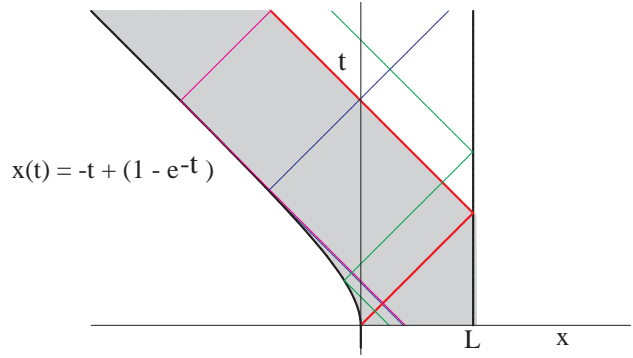,width=3.25in}
\end{center}
\caption[two mirrors]{A plot of null geodesics which contribute
to $g_0(x-t)$ at late times in a problem with
one moving mirror and one fixed mirror.  Note that eventually
the only contributions to fields which lie outside the shaded region
come from geodesics which reflect once from each
mirror or geodesics which reflect only from the moving mirror.}
\label{twomirror}
\end{figure}

Let us complete our discussion of flat-space problems by
considering a variant of the moving mirror in order to show that
for this problem, at late times, almost all of
space-time will be filled by fields that correspond to degrees of
freedom of the massless theory which were originally concentrated
in an exponentially small region of a single point.  This
peculiar, but otherwise absolutely trivial property of the massless
theory in the presence of a time dependent Hamiltonian,
will be encountered again when we discuss the question of
Hawking-Bekenstein entropy for the case of the black hole.  The
only modification we will make of the original moving mirror
problem will be to add a stationary mirror at
$x=L$, see Fig.\ref{twomirror}.

A glance at Fig.\ref{twomirror} shows that adding the second
mirror does not complicate the computation of the field operators
at any time in the future, since at most  two reflections should
be taken into account.  The interesting feature of this solution
to the field equations is that now, for times $t > L$ and points
$(t,x)$ lying to the right of the shaded region bounded by the
mirror and the line $x=2L-t$, there are no direct contributions to
the field operators.  Instead, the value of the fields at such a
point is the sum of contributions coming either from a single
reflection from the moving mirror or a reflection from the moving
mirror followed by a reflection from the stationary mirror. Thus,
we see that for large $t$ and points $(t,x)$ lying to the right of
the shaded region, the fields $\phi(t,x)$ and $\pi(t,x)$ only
depend upon the fields on the original $t=0$ surface which lie
within an exponentially small neighborhood of the point $x=A$.

\epsfverbosetrue

\section{The Black Hole: Geometric Optics Approximation}

Finally, let us turn to a  discussion of the Schwarzschild black
hole.  We pointed out in Section IV that by introducing
Lema\^itre coordinates we can canonically quantize the theory of a
massless scalar field and use the resulting time-dependent
Hamiltonian to derive the corresponding Heisenberg equations of
motion, Eq.(\ref{scheulerlag}).  As in the case of the
moving mirror, the behavior of the system at later times is
obtained by solving for the Heisenberg fields as a function of the
fields defined on the initial surface on which we quantized the
system.  In the flat space case we argued that the easiest way to
solve the Heisenberg equation of motion is to trace back the two
null-rays leaving  the point $(t,x)$ (the point at which we wish
to compute the field and its conjugate momentum), to the two
points from which they leave the $t=0$ surface and write the
answer in terms of the $\phi$ and $\pi$ at those two points. We
will now show how to generalize this approach to the corresponding
equations in curved space.

To motivate what we call the geometric optics approximation
let us study Eq.(\ref{scheulerlag}) in Painleve coordinates
($\lambda,r$), since these coordinates are non-singular and the
dependence of the solutions on $\lambda$ and $r$ factorizes. The usual WKB
approach to this sort of problem is to assume a solution of the form
$\phi_0 = r^{-1} e^{i \omega \lambda} f_\omega(r)$ and substitute
this ansatz into the field equation.  In this way one obtains that,
for large $\omega$, $f_\omega(r)$ can be written  as
\begin{eqnarray}
\ln f_\omega(r) &=& i\omega S_{1,2}(r) + {\cal O}(\omega^{-1}),
~~~
S_{1,2}(r) = \pm r -2\sqrt{r} \pm \ln((\sqrt{r}\pm 1)^2).
\label{solutions}
\end{eqnarray}
We now observe  that these
solutions are constant along incoming or outgoing null geodesics
where an incoming null-geodesic starting at the point $x_1$ at
time $\lambda=0$ is a curve $r(\lambda)$ such that
\begin{equation}
    S_1(x_1) = \lambda + S_1(r(\lambda)),
\label{ngone}
\end{equation}
and similarly, an outgoing geodesic starting at $x_2$ at $\lambda=0$, is
a curve $r(\lambda)$ such that
\begin{equation}
    S_2(x_2) =  \lambda + S_2(r(\lambda)),
\label{ngtwo}
\end{equation}
where $S_{1,2}$ are as defined in Eq.(\ref{solutions}).

To change this observation into an ansatz for the solution to the
S-wave field equation, we simply mimic the general form of the solution
for the moving mirror in flat space; i.e. we say that for general
$(\lambda,r)$
\begin{eqnarray}
    \phi_0(\lambda,r) &=& {1 \over r} \left( \tilde\phi_1 (\lambda + S_1(r))
    + \tilde\phi_2 ( \lambda+S_2(r) ) \right),
\label{geomopt}
\end{eqnarray}
and the functions $\tilde \phi_{1,2}(S_{1,2}(r))=f_{1,2}(r)$ are
to be determined from the boundary conditions
\begin{equation}
    \phi_0(0,r) = {\phi_1(r)\over r}, \quad
    \partial_\lambda \,\phi(\lambda,r)\vert_{\lambda=0} = \sqrt{r}\pi_1(r),
\label{bndrycond}
\end{equation}
where $\phi_1(r)$ and $\pi_1(r)$ are the rescaled operators we introduced
to quantize the theory on the initial surface $\lambda=0$.

Substituting Eq.(\ref{geomopt}) into Eq.(\ref{bndrycond}) and
following the same procedure as in the flat space case we obtain
\be
f_{1,2}(x) = \frac {1}{2}
\int \limits_{0}^{x} {\rm d} \xi \left [ \phi_1'(\xi)
\pm \pi_1(\xi) \mp \frac{ \phi_1(\xi)}{\xi^{3/2}} \right ],
\label{funcf}
\ee
where $\phi_1'= {\rm d}\phi_1/{\rm d}\xi$ and
$S_{1,2}(x_{1,2}) = \lambda + S_{1,2}(r)$.
Combined with the fact that field $\phi_1$ and its
momentum $\pi_1$ are expressed through the
creation and annihilation operators defined at $\lambda=0$,
the above set of equations  allows us to compute any Green's function
of the field $\phi_0$ at any later time.

Let us note that for the four-dimensional theory the geometric optics
form of the solution is not  valid as $r \to 0$. This
is a point we will return to when we discuss the information
paradox in the last section of this paper.

\section{Black Hole: Thermometer Redux}

As for the moving mirror, we expect the time dependence of the
Hamiltonian to reflect itself in the existence of steady state
phenomena such as outgoing radiation.  We will now show that this
is indeed what happens and that  this formalism naturally leads to
the prediction that there is an outgoing, time-independent flux of
energy which at large distances  corresponds to a body at a
Hawking temperature $T_{\rm H}$.  To do this we follow the
procedure used for the moving mirror and weakly couple the
massless field to a detector  \cite{unruh,dewitt79} (which acts as
a thermometer) located at some fixed Schwarzschild (or Painlev\'e)
radius $r$.  As before, we add an interaction term to the free
field Lagrangian of the form $V_{\rm int} \sim
e^{-(t-t_0)^2/(2\delta^2)} \phi_0(t,r) \hat Q$, where $Q$ is an
operator which acts in the Hilbert space of detector eigenstates.
Second order perturbation theory in $V_{\rm int}$ tells us that
the probability of exciting the detector to a state of energy $E$
is related to the Fourier transform of the Green's function of the
massless field \cite{birel}:
\ba
&& {\cal P}(\Delta E) \sim |\langle E | Q | E_0 \rangle |^2
\int {\rm d} t {\rm d} t' e^{-i\Delta E (t-t')
-[(t-t_0)^2+(t'-t_0)^2]/(2\delta^2)}
\langle  \phi_0(t,r) \phi_0(t',r)  \rangle,
\label{detex}
\ea
where $\Delta E = E-E_0$ and $E_0$ is the ground state energy of
the detector. The only difference is that now the Green's function
in Eq.(\ref{detex}) is to be computed using the evolution equation
for the field $\phi_0(\lambda,r)$ which relates it to $\phi_1(r)$
and $\pi_1(r)$  on the surface $\lambda=0$.

As in the case of the moving mirror it is convenient to define the points
$x_{1,2}$ and $y_{1,2}$ as follows:
\begin{eqnarray}
S_1(x_1) &=& \lambda_{1} +S_1(r_1),~~~
S_2(x_2) = \lambda_{1} +S_2(r_1), \nonumber\\
S_1(y_1) &=& \lambda_2 + S_1(r_2),~~~
S_2(y_2) = \lambda_2 + S_2(r_2),
\label{har}
\end{eqnarray}
so that  $x_1$ and $x_2$ are the
points on the $\lambda=0$ surface from which infalling and
outgoing null geodesics must leave to arrive at the point
$(\lambda_1,r_1)$ and $y_1$ and $y_2$ are the points from which
infalling and outgoing null geodesics must leave the same surface
to arrive at the point $(\lambda_2,r_2)$. Since we assume that the
thermometer stays fixed at the same Schwarzschild $r$ we must
identify $r_1=r_2=r$ and, also, we must remember that there is a
transformation between $\lambda$ and $t$ so that $\lambda_1 =
\lambda_1(t,r)$ and $\lambda_2 = \lambda_2(t',r)$.  At this point
we need to evaluate $x_{1,2}$ and $y_{1,2}$  for large
values of $t_0$.  Given the explicit form of the functions $S_{1,2}$
it follows directly that these equations have the approximate
solutions:
\begin{eqnarray}
x_1 &=& t +r,~~~
x_2 =  1 + 2  e^{-(t-r)/2 } , \nonumber\\
y_1 &=& t'+r,~~~
y_2 =  1 + 2  e^{-(t'-r)/2} ,
\end{eqnarray}
where we have assumed that $t \sim t' \sim t_0$.
Asymptotically, $x_2 \to y_2 \to 1$ and $x_1 \to y_1 \to \infty$.
In this limit, the Green's function can be written as
\ba
&& \langle \phi_0(t,r) \phi_0(t',r) \rangle \approx
\frac{-1}{4\pi r^2} \left (
\ln|x_1 - y_1| + \ln|x_2 - y_2|
 + \frac{i \pi}{2}
\left [
\kappa (x_1,y_1) + \kappa (y_2,x_2)
\right ]
+ c
\right ),
\label{gf}
\ea
where $\kappa(x,y) = \theta(x-y) - \theta(y-x) $
and $c$ is some constant.

It is instructive to consider the terms in Eq.(\ref{gf})
separately. The constant does not contribute to ${\cal P}(E)$
since, as in the case of the moving mirror, it yields a result
proportional to $\exp(-\Delta E^2 \delta^2) \ll 1$. The $\ln|x_1 -
y_1|$ term and the terms described by the function
$\kappa$ give simple contributions that can be written as
\be
{\cal P}_1(\Delta E) \sim - \frac{\pi \delta}{\Delta E}
+ {\cal O}((\delta \Delta E)^{-1}).
\label{c1}
\ee
The important part of the final result comes from the second term
in Eq.(\ref{gf}) which describes the radiation coming from the
vicinity of the horizon. Appropriately shifting the integration
variables and restoring factors of $2M$ we obtain
\ba
{\cal P}_2(\Delta E) && \sim
- \int {\rm d} t {\rm d} t' e^{-i\Delta E (t-t') }
e^{-[(t-t_0)^2+(t'-t_0)^2]/(2\delta^2)}
\ln|e^{-t/(2M)}-e^{-t'/(2M)}|.
\ea
If we then change the variables to $v=(t+t'),~u=(t-t')$ and neglect
all the suppressed terms we arrive at
\be
{\cal P}_2(\Delta E)\sim \frac{2\pi \delta}{\Delta E}
\left [\frac{1}{e^{\Delta E/T_{\rm H}} -1} + \frac{1}{2} \right ],
\ee
where the Hawking temperature $T_{\rm H} = 1/(8\pi M)$ has been introduced.
Since the total probability is given by the sum of ${\cal P}_1$ and ${\cal P}_2$
we have the final result:
\ba
\frac{{\cal P}(\Delta E)}{\delta} \sim
\frac{|\langle E | Q | E_0 \rangle |^2 }{\Delta E}
\times
\frac{1}{e^{\Delta E/T_{\rm H}} -1}.
\ea
The interpretation of this formula is straightforward. If, at a
large, fixed distance from the black hole, an observer switches on a
detector which interacts with the massless field for finite amount
of time, then the energy levels of the detector get populated as
if the detector was in equilibrium with a thermal distribution of
particles at a temperature $T_{\rm H}=1/8\pi M$.

The only subtlety which we have skipped over in this calculation
done for large $r$, is that for arbitrary $r$ the interaction term
should have a correction for the time dilation at point $r$, since
the energy levels of the thermometer are defined in its rest
frame.  The result of adding this to the calculation is that, putting
back factors of $M$, a thermometer at arbitrary $r$ will read a temperature
\begin{equation}
k_B\, T = {1 \over 8 \pi M \sqrt{1 - 2M/r}} .
\end{equation}

\section{Black Hole: Energy Flux}

The Schwarzschild calculation of the energy flux through a sphere
of fixed radius is done in the same way as for the moving mirror:
we point-split the fields appearing in the expression for
the energy momentum tensor, regulate the resulting expression and
then take the limit of zero splitting.  The result of this computation
is that we find that $T_{\lambda,\eta}$ is finite and non-vanishing and
the total flux through a sphere of large radius is given by
\begin{equation}
  {\rm Flux} = {\pi \over 12}\ {T_H}^2  = {\pi\over 12}\ {1 \over (8 \pi M)^2} .
\end{equation}
The full expression for $T_{\lambda, \eta}$ contains terms which
vanish for large $\lambda$ and therefore can
be identified as transients;
persistent terms which decrease faster than $1/r^2$ and, finally,
persistent terms which fall off as $1/r^2$ and hence contribute
to the flux. While we have carried out the computation for arbitrary
$(\lambda,r)$, the resulting expressions are too cumbersome to
present here and we limit our discussion to the computation of
terms that both approach  a constant for large $\lambda$ and fall off
as $1/r^2$.

Let $\sigma = \lambda_2 - \lambda_1$ and define the flux as:
\be
\langle T_{\lambda \eta} \rangle = \lim_{\sigma \to 0} \frac{1}{2}
\langle 0 |
\left \{ \frac{\partial \phi_0 (\lambda_1,\eta)}{\partial \lambda },
\frac{\partial \phi_0(\lambda_2,\eta) }{\partial \eta}  \right \} |0 \rangle.
\label{fluxdef}
\ee
The formula for the total energy passing through a large sphere,
in Lema\^itre coordinates, is given by
\be
{\cal J} = \lim_{\eta \to \infty} \int {\rm d} \phi {\rm d} \theta
\sqrt{-g}  g^{\lambda \lambda} g^{\eta \eta}
\frac{\langle T_{\lambda \eta} \rangle}{4\pi}
 =-\lim_{\eta \to \infty} \frac{r^{5/2}}{(2M)^{1/2}}
\langle T_{\lambda \eta} \rangle,
\label{flux}
\ee
where a normalization factor of $(4\pi)^{-1}$ is
introduced since $\phi_0$ denotes the $S$-wave component of the
massless field. Given this expression, it is straightforward
to compute this flux using the explicit expression for the
time evolution of the field $\phi_0$, Eq.(\ref{geomopt}), and then
take the limit
$\lambda_2 \to \lambda_1$.

Since in the geometric optics approximation the field and
its conjugate momentum are given in purely geometrical terms
(i.e., in terms of the point at which the field is to be
evaluated, the null geodesics arriving at that point and the two
points on the initial surface of quantization from which they
left), we are free to carry out the computation in any
coordinate system. Since the calculation of the flux for large
values of $\lambda$ and $\eta$ is simplest in Painleve coordinates
we will transform to these coordinates in the discussion which
follows.  However, because we define the point-split energy
momentum tensor by holding $\eta$ fixed and separating the fields in the time
variable $\lambda$ we have to take into account that the
variables $r_1$ and $r_2$ corresponding to $(\lambda_1,\eta)$ and
$(\lambda_2,\eta)$ are related by
\be
r_2 = r_1 - \frac{\sqrt{2M}\sigma}{r^{1/2}}
-\frac{2M \sigma^2}{4r_1^2}
-\frac{1}{6} \frac{(2M)^{3/2} \sigma^3}{r_1^{7/2}} + {\cal O}(\sigma^4).
\ee
Taking the derivatives in Eq.(\ref{fluxdef}) and considering
the limit $r_{1,2} \to \infty$, we obtain the following
expression for the flux
\ba
&& {\cal J}= -\frac{1}{2} \lim_{\sigma \to 0}
\left [ \langle \left \{
f_1'(x_1),f_1'(y_1)
\right \} \rangle
W_1(r_2,r_1,x_1,y_1)
+
\langle \left \{
f_2'(x_2),f_2'(y_2)
\right \} \rangle
W_2(r_2,r_1,x_2,y_2)
\right ],\nonumber\\
\label{e16}
\ea
where the functions $f_{1,2}$ are defined in Eq.(\ref{funcf}), and
\begin{equation}
W_i(r_2,r_1,x,y) = \frac{S_i'(r_2)}{S_i'(x)S_i'(y)}
\left (1 - \frac{S_i'(r_1)}{\sqrt{r_1}} \right).
\end{equation}
Once again (cf.  Eq.(\ref{har})) we define $x_{1,2}$, $y_{1,2}$
to be the points on the $\lambda=0$ surface from which the
null-geodesics which wind up at the point $(\lambda,\eta)$ originate.
Given these equations it is easy to derive the relation
between $y_{1,2}$ and $x_{1,2}$ as a power series expansion in
$\sigma$. It follows from the specific form of the functions
$S_{1,2}$ that, in the limit of large $\lambda$ and large $\eta$,
the limiting values for these points are $x_1 \to \infty$, $x_2/2M
\to 1$.

To compute the expectation value of the anticommutator $\langle
\{f_1'(x_1),f_1'(y_1)\} \rangle$ we start from Eq.(\ref{funcf}) and
derive
\ba
 \langle \{f_1'(x_1),f_1'(y_1)\} \rangle
=  \frac{1}{4} \left [
\langle
\{\pi_1(x_1),\pi_1(y_1) \}\rangle
 + \langle \left \{\phi_1'(x_1), \phi_1'(y_1)
\right\}\rangle +...\right], \nonumber
\ea
where the dots represent terms which do not contribute to the flux in the
$r_1 \to \infty$ limit and the anticommutators are
computed using Eq.(\ref{ffexpansion}).  In this way we find
\ba &&
\langle \left \{\phi_1'(x_1), \phi_1'(y_1) \right\} \rangle =
 \frac {-2}{\pi} \frac {(y_1^2+x_1^2)}{(x_1^2-y_1^2)^2},
~~~~\langle \{\pi_1(x_1),\pi_1(y_1) \} \rangle  =
- \frac{4x_1 y_1}{\pi (x_1^2 - y_1^2)^2},
\ea
so that the final result for the anticommutator reads
\be
\langle \{f_1'(x_1),f_1'(y_1)\} \rangle
 = -\frac{1}{2 \pi} \frac{1}{(x_1 - y_1)^2} + ...
\ee
Performing a similar calculation for the second term in Eq.(\ref{e16}) and
substituting expansions of $y_{1,2}$ and $r_2$ in terms of $x_{1,2}$ and
$r_1$, we find for the energy flux
\be
{\cal J} = \frac{1}{192 \pi (2M)^2} = \frac{\pi}{12} T_{\rm H}^2,
\label{fluxres}
\ee
where once again we have introduced the Hawking temperature $T_{\rm H} = 1/(8\pi M)$.

Eq.(\ref{fluxres})  shows that the
energy flux at large distances is finite.  We have already noted that
this result is in accord with a general theorem that deals with the
structure of the possible divergences in the stress-energy tensor
computed in a gravitational background  \cite{birel}.
It turns out  that one can derive an explicit formula
for all of the possible divergences which can occur as
coefficients of specific functions of the metric and renormalize
them by adding explicit counterterms to the Einstein Lagrangian.
Adding  these terms to the Lagrangian and computing  the
resulting modifications of the energy momentum tensor, one finds
that there are no terms which can contribute to the {\it off-diagonal}
element of the energy momentum tensor in both the  flat
space and the Schwarzschild metric in  Lema\^itre coordinates.
Thus, since there are no possible counterterms which can remove
divergences in the flux,
the result must come out finite as  it indeed does.
Unfortunately, for finite values of $r$
our result still contains logarithmically divergent terms
${\cal O}(\ln \sigma)$ multiplied by functions of $r$ that
decrease faster than $1/r^2$ in the limit of large $r$.
We believe this to be due to the fact that
we have restricted attention to the $L=0$ component of the field $\phi$
and have not considered higher angular momenta.
This observation is supported by the fact that in the case of a
two-dimensional black hole, where higher angular momentum modes
are absent, our result for the flux is finite for all values of
$r$.

To conclude our discussion of the flux let us comment a bit on the
back reaction issue. The problem of back reaction is equivalent
to the statement that the computation we are doing is not, from the
point of view of the Einstein equations, self-consistent even at
the semi-classical level since, on the one hand,  for a static Schwarzschild
metric the Einstein tensor $G_{\mu\nu}$ vanishes for all $r \ne 0$
but, on the other hand,
we find that the energy-momentum tensor of the scalar field has, at
large times, a finite, uniquely defined, off-diagonal component in
this background.

Since  our approach studies the behavior of the scalar field
theory starting from a well defined quantum state at a
finite initial time, we should be able to  solve the problem iteratively,
computing corrections to the Schwarzschild metric due to non-zero
expectation value of the stress-energy tensor at the right hand side
of the Einstein equations. Although it is quite difficult to do this
in general, we would like to point out that a simple modification
of the metric:
\begin{eqnarray}
    ds^2 = -(1-{2M(t)\over r})\,dt^2 + {1\over (1- {2M(t)\over r})}\,dr^2
    + r^2 d\Omega^2.
\label{tSchwz}
\end{eqnarray}
produces the Einstein tensor of the form
\begin{equation}
\frac{-2}{X(t) r^2} \frac{{\rm d}M(t)}{{\rm d}t}
\pmatrix{
0 & 1 &0 & 0 \cr
1 & 0 & 0 & 0 \cr
0 & 0 & 0 & 0 \cr
0 & 0 & 0 & 0 \cr
}
\label{eintensor}
\end{equation}
where $X(t) = 1-2M(t)/r$ and where we only retained
terms which are linear in ${\rm d} M(t) / {\rm d} t$.

Although the approximation for the metric Eq.(\ref{tSchwz}) is
too crude to reproduce the vacuum expectation value of the
stress energy tensor of the massless field for arbitrary values
of $t$ and $r$, for large values of $r$ we can easily match
the expression for the Einstein tensor, Eq.(\ref{eintensor}), to
the vacuum expectation value for the energy flux.
We then obtain the well known equation that describes the evaporation
of the black hole:
\be
 \frac{{\rm d}M(t)}{{\rm d}t}
= -\frac{\pi}{12} \left  ( \frac{1}{8\pi M} \right )^2.
\ee

In principle, it should be possible to iteratively improve on this
approximation and in this case solve the back reaction problem
at the quasi-classical level until relatively late in evaporation process,
thus obtaining a better insight into the black hole dynamics.

\section{An Infalling Reflecting Mirror}

We have now discussed the canonical quantization of a massless scalar
field in the background of an eternal Schwarzschild black hole and have
shown that, in the geometric optics approximation, one can compute
all of the usual results such as the Hawking temperature, as
measured by a thermometer held at fixed Schwarzschild $r$, and the
flux through a sphere of the same radius. Now we would like to
address the question of the choice of initial state on the surface
of quantization.  We have already demonstrated that on any
fixed Painlev\'e or Lema\^itre time-slice the Hamiltonian exhibits
no singularities and, in fact, is the Hamiltonian of an ordinary
massless free field theory.  Thus, there is no problem in
defining the $\lambda=\lambda_0$
Hilbert space.

The only issue which remains is a particular choice of the initial
state for the field theory on the quantization surface. At this
point we would like to stress that practically for any choice of
the initial state at $\lambda=0$ surface which differs from the
vacuum state by a finite amount of energy, the large time behavior
of the system will be exactly the same as described in the
previous Sections.  However, we should also note that a sufficiently
bizarre choice of the initial state can result in no Hawking
radiation at late times.

To show that such choices of the initial state are unnatural, we
would  like to pose a problem which uses a stationary reflecting
mirror to guarantee that for infinite times in the past there is a
natural choice of initial state.  Next, we let the mirror start to
fall into the Schwarzschild black hole along a Lema\^itre timeline
and ask if, starting from this well defined initial state, we see
Hawking radiation at late times.  We will argue that we do and
that this easily follows from  the formalism that we have already
discussed in detail.

\epsfverbosetrue
\begin{figure}
\begin{center}
\leavevmode
\epsfig{file=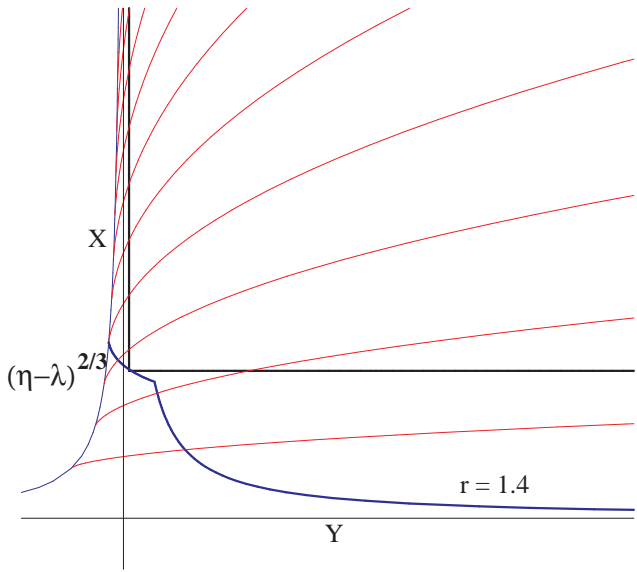,width=3.25in}
\end{center}
\caption[InfallingMirror]{The curved dark line represents a mirror
which initially, for infinite times in the past, sits at Schwarzschild $r=1.4$
and then starts to fall into the horizon along a trajectory $r(\lambda,\eta)$.
The horizontal and vertical lines represent an incoming null-geodesic hitting
the infalling mirror just before it crosses the horizon and being reflected
outwards. From the picture it is clear that the closer it is to the horizon
when it is reflected, the longer it takes to get out to large Schwarschild  $r$.}
\label{infallingmirror}
\end{figure}

The problem we have in mind is a generalization of the problem of
the moving mirror to the case in which the mirror moves in the
background of a Schwarzschild black hole.  To be specific we assume,
as is shown in Fig.\ref{infallingmirror}, that we are dealing with
a perfectly reflecting spherical mirror (i.e. a surface on which
we assume the scalar field must vanish) of Schwarzschild radius $R
\gg 1$. Furthermore, as with the moving mirror we assume that the
field $\phi(r)$ exists only in the region outside the mirror.
Next, we assume that at some finite Painlev\'e time $\lambda_1$
this mirror starts to fall in along one of the Lema\^itre curves
$r(\lambda) = (R^{3/2}-3(\lambda-\lambda_1)/2)^{2/3}$. In the
geometric optics approximation this problem is solved in
essentially the same way as in our earlier discussions, except now
we have to take into account contributions coming from null
geodesics which fall in towards the mirror and then are reflected
out.  If we quantize on an initial surface $\lambda_0 \ll
\lambda_1$, then for all times less than $\lambda_1$ the field in
the region $r \ge R$ comes from geodesics which propagate in,
essentially, flat space.  Comparing this with our previous
calculation we see that for this region there is no generation of
Hawking radiation.  Remember, that in order for the thermometer to
measure a Hawking temperature, or for the energy momentum tensor
to show a non-vanishing flux of apparently thermal radiation, two
events which are separated by a small interval in $\lambda$ must,
when traced back to the initial surface of quantization by null
geodesics, come from points which are exponentially closer
together than when they started out.  We saw in the previous
discussion that this only happens when these null geodesics pass
very close to the horizon.  This means that the radiation comes
from the rays which are reflected from the mirror
as it passes through an exponentially small region near the
horizon (see  Fig.\ref{infallingmirror}).
From this argument we see that
the Hawking radiation seen at large future times is generated from
points on the initial surface of quantization which all lie in a
small neighborhood of the point $r_0$, defined by the condition
that an infalling null geodesic drawn from this point would hit
the mirror just as it passed through the horizon.  This point
$r_0$ plays essentially the same role as the corresponding point
$x_0(t,x)$ in our original discussion of the moving mirror.

Therefore, in order to suppress the Hawking radiation in this
setting, one would need to significantly modify the vacuum
expectation values of field operators and their products at what
is  essentially an arbitrary point, $r_0$.  It is this strange
requirement on the initial state that we consider to be
unnatural and from this we conclude that the Hawking radiation is
a robust phenomenon practically independent of the initial state.

\section{Entropy}

The next topic we would like to discuss is the question of the
entropy of the black hole.  Clearly if we deal with a Hamiltonian
system that starts in a pure state and experiences completely
unitary evolution it has zero entropy (since the entropy of a pure
state is zero) despite the fact that it exhibits all of the
phenomena associated with Hawking radiation.  Nevertheless, despite
this obvious statement we can slightly modify the problem we
have been discussing and construct an object which, from the point
of an outside observer appears to have an energy $M$, temperature
$T_H = 1/8\pi M$ and an entropy $S= A_H/4$, where $A_H$ is the
area of the horizon, $A_H = 4\pi M^2$.  Let us see how this
is done and then ask what is happening.

For this purpose consider a  black hole surrounded by a
perfectly reflecting mirror of a large radius $R$.
In contrast to the infalling mirror case, however, we
now assume that the field degrees of freedom live inside the
spherical mirror; i.e., in the region $0 \le r \le R$. By putting
this mirror around the hole we don't allow any energy to escape
the region surrounded by the mirror, since $T_{\mu \nu}$ is
locally conserved; hence the total energy of the enclosed region is,
for all times, given by the original mass of the black hole $M$,
which can be measured by an outside observer, simply
by dropping a test particle and measuring its acceleration.
Next, imagine that the outside observer has a thermometer inserted
through a very small hole in the mirror.  Clearly, nothing is
changed in the calculation presented previously and so this thermometer
measures a Hawking temperature $T_H$.  Thus, we see that we have
a macroscopic object which appears to have an equation of state
that says the energy of the object is inversely proportional to its
temperature.  If we now try to use thermodynamic concepts, the usual
formula $dU = T\,dS$ becomes
\begin{equation}
    dM = {1 \over 8\pi M}\,dS,
\end{equation}
so that by integrating it and by setting the entropy of a zero mass object
of this type to  zero, one derives:
\begin{equation}
    S = 4\pi M^2 = {A_H \over 4}.
\end{equation}
This is the usual Bekenstein argument.  Note however that the
entire discussion we have just given is being applied to a quantum
system during the period when it is certainly described by a pure state.
This makes it problematic to assign a non-vanishing entropy
to the system and so are we are left asking the question
what the entropy of the system constructed in such a way means
and what it tells us about the system.

In contrast to the standard notion of entropy in the equilibrium
thermodynamics, the entropy of the system black hole plus
the massless field does not tell us anything important since
we must only look inside our reflecting mirror and note
that the system whose properties are being measured
is never in equilibrium.  This lack
of equilibrium is not because the black hole is evaporating, but
because the apparently thermal flux arriving at the mirror comes
from the horizon, is reflected and then disappears through
the horizon to be stored at $r=0$.  What we are looking at is a
steady state phenomenon in which an exponentially small region
near the horizon serves as a constant source of new radiation with
essentially the same properties as the reflected radiation which
then disappears behind the horizon. At first these assertions
might seem peculiar, but we would remind the reader that this sort
of phenomenon was already seen in the case of the two-mirror
problem. Another feature of the two-mirror problem which is
intimately related to this behavior is that the two null geodesics
attaching to the point $(\lambda,r)$, for large enough $\lambda$,
either come directly from an exponentially thin region around the
horizon, or are rays which originated from such a region at an
earlier time and then were reflected back to arrive at
$(\lambda,r)$.  Hence, in the geometric optics approximation, we
see that for sufficiently large $\lambda$ the fields between the
horizon and the mirror are functions of fields which lay within an
exponentially small region around the horizon on the initial
surface of quantization. This phenomenon is the source of both the
Hawking radiation and the reason why late time Green's functions
seem to show no correlations over finite separations in time and
space. From this argument we see that the Bekenstein entropy is
somehow related to the curious property seen in the problem of the
moving mirror, that the causal structure of the theory guarantees
that after a finite time the fields between the horizon and
reflecting shell are only functions of the degrees of freedom
localized within an exponentially thin shell surrounding the
Schwarzschild radius.  At best, for this problem, the
Hawking-Bekenstein entropy is a reflection of this fact.

\section{The Information Paradox}

Given that a Hamiltonian formulation of the problem of a massless
scalar field in the background of a large Schwarzschild black hole
exists, there cannot be an information paradox until one comes to
grips with the question of what happens during the final moments
as the black hole evaporates.  While our approach doesn't allow us
to discuss these violent final moments of the process of black hole
evaporation, it does provide insight into the question
of where and how degrees of freedom which "fall into the black
hole" are stored.  It also provides a different picture as to what
might happen after evaporation has taken place. It is this picture
of the evolution of the problem from its initial state to late
times that we discuss in this section.

As noted earlier, in the case of a four-dimensional black hole the
"geometric optics" approximation fails to give an accurate description
of what is happening near $r=0$  but,  fortunately, this failure
of the approximation is not an insuperable obstacle to obtaining a
more complete understanding of the physics.  There are two reasons
for this.  First, if we restrict attention to the case of a
two-dimensional black hole (i.e., the theory obtained if we
restrict the metric to just the upper $2\times 2$ matrix $g_{\mu
\nu}$, $\mu,\nu=0,1$), the geometric optics approximation is in
fact exact.  Second, there exists a useful discretization of the
problem in Lema\^itre coordinates which allows one to consistently
investigate the problem in both two and four dimensions.

The importance of the fact that the geometric optics approximation
is exact for the case of the two dimensional black hole is that it
tells us what is happening at $r=0$.  Our treatment begins by
canonically quantizing the massless scalar field theory on a
surface of constant Painlev\'e time, $\lambda = \lambda_0$, and so
we are free to require that on this surface the field vanishes at
$r=0$.   For subsequent times, however, this boundary condition
will be true if and only if it is consistent with the Heisenberg
equations of motion and a simple computation shows that for
$\lambda > \lambda_0$ the field does not vanish.

To understand the context of this observation, let us note that
the line $(\lambda,0)$ is spacelike.  Thus, what we
learn from the solution to the field equations is that  when we
restricted the integration over $\eta$ to run from $\lambda \le
\eta \le \infty $, in Eq.(\ref{hamone}), we were not doing the correct
thing; we should have included the spacelike surface $r=0$ running
from $\lambda_0$ to $\lambda$.  This prescription should have been
obvious from the situation shown in Fig.\ref{lemaitretime},
where we see that as $\lambda$ increases the surface of fixed
Painlev\'e time $\lambda_0$ gets mapped onto the line $r=0$ and
the surface of fixed Painlev\'e time $\lambda$. Given that we know
this prescription is required in two dimensions it is not much of
a stretch to assume that the same is true in the four dimensional case.

Because, in four dimensions, the geometric optics approximation
is not valid near $r=0$ we need to do something else to get a
better feeling for the problem, something which agrees with the
geometric optics approximation in two dimensions.  To arrive at
such a treatment of the problem we need to deal with two issues.
The first issue has to do with the fact that $r=0$ is
the location of a true singularity in the metric, where the
curvature diverges and one expects
quantum gravity to play a role. Thus a full treatment
of this problem would perforce need to go beyond the semi-classical
approximation.  A less profound technical problem is that
the Hamiltonian treatment along the line $r=0$ needs to be
carefully done, especially at the points
where the line $r=0$ and the
surface of constant Painlev\'e time meet.  While we have nothing
to say about what the true quantum completion of the theory of
gravity might be, the second problem is easily handled if we
assume a minimum value for $r_{\rm min}=\epsilon$,
formulate the Hamiltonian problem and then take the limit
$\epsilon \rightarrow 0$.  For the two-dimensional problem this
amounts to quantizing the field theory on the surface of Painlev\'e
time $\lambda_0$ imposing the condition that the field vanishes at
$r=\epsilon$ and then using the geometric optics construction to
evaluate the field and its conjugate momentum along the curve
$(\lambda,\epsilon)$ in the future.  Unfortunately, this simple
construction is not sufficient to handle the four dimensional theory.

In order to study both the two and four dimensional problems
in a consistent manner, we propose to discretize the problem
by introducing a lattice in $\eta$. To do that we replace the continuous
variable $\eta$ by a discrete variable $\eta_j$ defined by
\begin{equation}
    \eta_j = \delta\,(\epsilon + j)
\end{equation}
where $\delta$ is a lattice spacing which has dimensions of
$(length)^{3/2}$ and $\epsilon$ is a dimensionless parameter
such that $r_{\rm min}=(3 \delta \epsilon/2)^{2/3}$.
With this definition of the lattice we then introduce dimensionless
rescaled fields $\Phi$, $\Pi$ and the rescaled time variable
\begin{eqnarray}
    \Phi(\lambda,\eta) &=& \delta^{2/3} \phi(\lambda,\eta),
~~\Pi(\lambda,\eta) = \delta^{1/3} \pi(\lambda,\eta),
~~\lambda \rightarrow {\lambda \over \delta^{2/3}}
\end{eqnarray}
and rewrite the time dependent Hamiltonian as:
\begin{eqnarray}
\label{hoflambda}
H(\lambda) = {1 \over \delta^{2/3}}
\sum_{j=1}^\infty \left [{2 \Pi(\lambda,j)^2 \over 3\eta(\lambda,j)}
+  \left ( {3\over 2} \right )^{5/3} \eta(\lambda,j)^{5/3}
(\phi(\lambda,j+1)-\phi(\lambda,j))^2
\right ],
\end{eqnarray}
using an additional assumption that
\begin{equation}
    \eta(\lambda,j) = \epsilon\, \theta(\lambda-\epsilon-j)
    + (\epsilon + j- \lambda)\,\theta(\epsilon + j - \lambda) .
\end{equation}
This is in accord with our previous remarks which say that once
the field reaches the point $\eta(\lambda) = \delta(\epsilon + j
- \lambda) = \delta\epsilon $ it stays there.

At this point we should make two observations. First, it is
obvious that the number of degrees of freedom remain the same in
this latticized version of the massless field theory and the
entire effect of the metric appears in the time dependence of the
coefficients appearing in the Hamiltonian.  The second, is that
the lattice we have introduced is somewhat peculiar. This is
because spacing lattice sites equally in the variable $\eta$ does
not correspond to spacing them equally in Schwarzschild $r$.  In
fact, since $r(\lambda) = (3(\eta-\lambda)/2)^{2/3}$ we see that
for large values of $r$ the spacing between two neighboring
lattice points decreases like $\delta/\sqrt{r}$.  Thus, while the
lattice provides a good cutoff for the field theory inside the
Schwarzschild radius it is not an effective cutoff for the theory
at large $r$ and the discussion we gave for continuum theory at
large $t$ and $r$ continues to be necessary.  Since, for now, we
are most interested in the behavior of the theory near $r=0$, this
latticized version of the field theory is what we need in order to
discuss the physics in a way which goes beyond the geometric
optics approximation.  Because the lattice Hamiltonian we have
introduced is explicitly time dependent, a full discussion of the
physics of the theory would require computing the unitary time
development operator $U(t)$, which is beyond the scope of the
present paper. Despite this, we can use this formulation of the
theory to gain some better understanding of what is happening and
to discover a peculiar property of the combined system of the
massless scalar field in a Schwarzschild background.

The first question which we can address using this formalism
is whether the apparent time dependence of the problem is a coordinate
artifact, in that we have chosen a coordinate system which, although
free of singularities, makes the metric time dependent.  Fortunately, we
can buttress our claim that the problem is intrinsically time dependent,
even without computing the operator $U(t)$, by observing that
the spectra of the instantaneous Hamiltonians $H(\lambda)$ are
changing as a function of time.  Qualitatively we can see that this is
the case by comparing the spectrum of the Hamiltonian, Eq.(\ref{hoflambda}),
for $\lambda=0$, against what it would be for a relatively large
value of $\lambda$.  When $\lambda=0$ and $\delta$ is very small
this Hamiltonian will be a discrete version of the continuum Hamiltonian
we discussed earlier and we expect the spectrum to be that of the
zero angular momentum mode of a free massless field theory, i.e.,
proportional to $k^2$.  Now, when $\lambda$ is large, then approximately
$\lambda$ lattice sites lie on the curve $r=\epsilon$
and the part of the Hamiltonian $H(\lambda)$ which refers only to
these sites has the form
\begin{equation}
H_1 = \sum_{j=0}^{\lambda} {2 \Pi(j)^2 \over 3\epsilon} +
({3\over 2} \epsilon)^{5/3} \,(\phi(\lambda,j+1)-\phi(\lambda,j))^2
\end{equation}
which after  rescaling  $\Pi(j)$ to absorb the factors of $\epsilon$,
becomes
\begin{equation}
H_1^{'} = \sum_{j=0}^{\lambda} {\Pi^{'2}\over 2} + {1\over 2}\,
4 \left ({3\epsilon \over 2} \right )^{2/3}\,
(\phi^{'}(\lambda,j+1)-\phi^{'}(\lambda,j))^2
\end{equation}
Since this has the form of a latticized nearest neighbor interaction we
see that for large $\lambda$ the spectrum of the kinetic term will be
\begin{equation}
{\cal E}(k) = 4 \left ({3\epsilon \over 2} \right )^{2/3}(1 - \cos(k))
\end{equation}
where $0 \le k \le 2\pi/\lambda$ since $\lambda$ is the number of
sites on the lattice.  The remaining part of the Hamiltonian
consists of two pieces.  The first piece, which we will ignore for
now, is a single term linking the curve $r=\epsilon$ (characterized by
the condition that $\eta(\lambda,j)=\epsilon$) and the points for
which $\eta(\lambda,j)=\epsilon+j-\lambda$, and the second piece that
consists of an infinite number of terms having essentially exactly
the same form as the Hamiltonian for $\lambda=0$ which has a
spectrum which goes like $k^2$.  Clearly the growth of a part of
the spectrum which behaves like a free field with a very different
speed of light from the rest of the theory represents a major
change in the spectrum and thus constitutes a proof that this set
of Hamiltonians really represent time dependent physics.

\begin{figure}
\begin{center}
\leavevmode
\epsfig{file=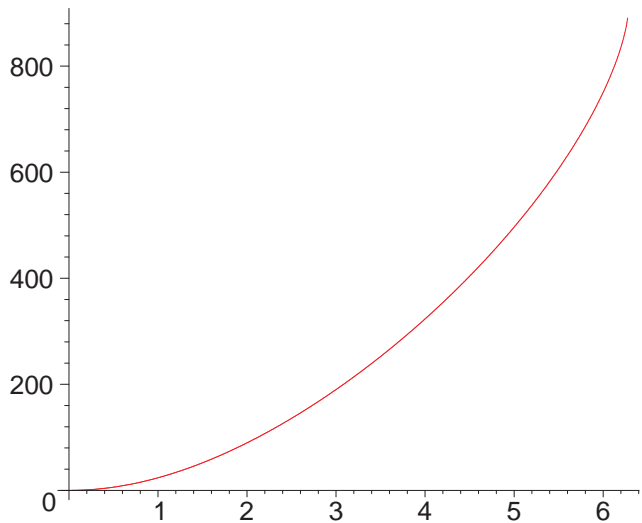,width=3.25in}
\end{center}
\caption[latticeonecap]{A plot of the eigenvalues of the latticized
Hamiltonian for 800 lattice sites with $\epsilon=0.001$.  Energies
are plotted versus the variable $k_p = 2\pi p/800$, for $p=0..799$.}
\label{latticeone}
\end{figure}

\begin{figure}
\begin{center}
\leavevmode
\epsfig{file=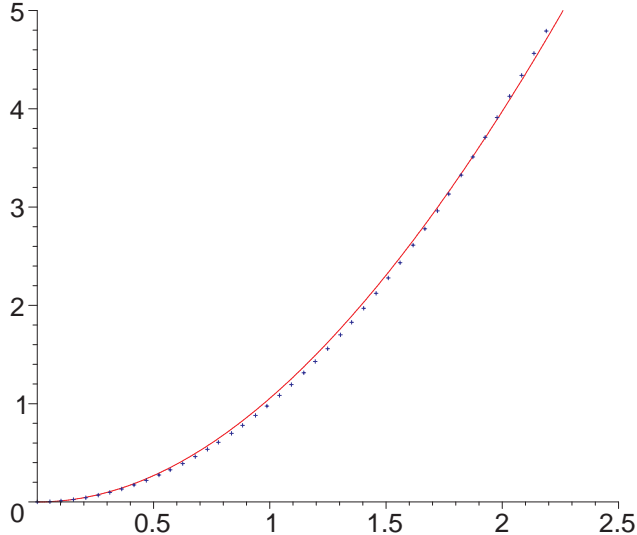,width=3.25in}
\end{center}
\caption[energyvsksqcap]{We have scaled the energy to run from $0$ to $2\pi^2/800$
vs square of momentum for $k_p = 2\pi p/800$.  The plot is limited to
a bit less than half of the total range to show how small the deviations
are from what we would expect in the infinite volume limit.}
\label{energyvsksq}
\end{figure}

In an attempt to achieve a better understanding of how these two
different free field theories are linked together, we diagonalized
the Hamiltonian in Eq.(\ref{hoflambda}) for lattices which initially,
in Schwarzschild coordinates, cover the region $0 \le r \le 20
R_S$, where $R_S$ stands for the Schwarzschild radius.
Fig.\ref{latticeone} shows a plot of the spectrum for a lattice
with 800 lattice sites, $\delta = 0.025$ and $\epsilon = 0.001$.
The energy is plotted against the variable $k_p = {2\pi p \over
800}$ where $p$ is an integer running from $0$ to $709$ and
represents what the lattice momentum would be for a very large
number of lattice points.  In this case we expect small
corrections due to the finite size of the matrix corresponding to
the kinetic term.  We already noted that for a very large lattice
we expect this plot to be proportional to $k^2$.  In
Fig.\ref{energyvsksq} we plot this result scaled to have a maximum
value of $(2\pi)^2$ against $k^2$ over almost half the range of
$k$ in order to show that even for a finite lattice the agreement
with our expectations is quite good.

\begin{figure}
\begin{center}
\leavevmode
\epsfig{file=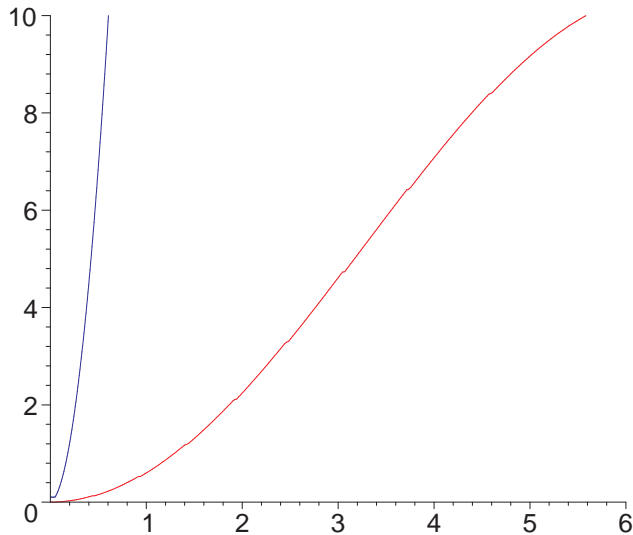,width=3.25in}
\end{center}
\caption[latticetwo]{The lower curve represents a plot of the first
300 eigenvalues of the Hamiltonian for $\lambda = 300$.  These correspond
to the number of points which lie on $r=\epsilon$ at this time.  This
curve is scaled to make it visible and it is plotted vs momenta $k_p = 2\pi p/300$.
The second curve represents the remaining eigenvalues, which naively correspond
to the fields lying between $r=\epsilon$ and $r=20 R$.  This is plotted vs the
variable $k_p = 2\pi p/900$.}
\label{latticetwo}
\end{figure}

Next, in order to exhibit the difference between this situation and
that for finite $\lambda$, we plot the eigenvalue spectrum of the
Hamiltonian for a lattice of $1200$ sites, $\epsilon=0.001$,
$\delta = 0.016$ and $\lambda = 300$.  This is a situation
which corresponds to having about $300$ lattice sites on the
curve $r=\epsilon$ and $900$ points extending from $r=\epsilon+1$
out to a distance of $20$ times the Schwarzschild radius.
Fig.\ref{latticetwo} shows two curves.  The first is a plot
of the lowest $300$ eigenvalues of the kinetic term divided by
$\epsilon^{2/3}$ and plotted against a momentum variable $k_p = 2\pi p/300$,
where we have scaled the eigenvalues to make them show
up on the plot.  The choice of momentum variable is motivated by
the fact that, as we have noted, the first $300$ terms in the
Hamiltonian would have a spectrum almost proportional to
$(1-\cos(k_p))$ if we dropped the term linking them to the next
$900$ terms.  The second curve is a plot of the next $900$
eigenvalues of the Hamiltonian versus a momentum $k_p = 2\pi/900$.
This should be quite similar to the curve shown in
Fig.\ref{latticeone} and it is.  We should emphasize that while for the case
$\lambda=0$ this spectrum starts at zero, in the case
$\lambda=300$ it doesn't; admittedly that fact is
not obvious in this plot.

\begin{figure}
\begin{center}
\leavevmode
\epsfig{file=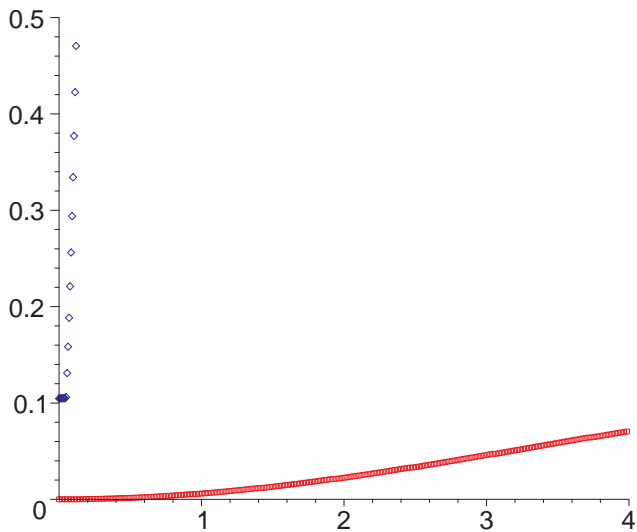,width=3.25in}
\end{center}
\caption[latticethree]{A plot of the two pieces of the eigenvalue
spectrum of the $\lambda=300$ Hamiltonian on an expanded
scale to show that the $r=\epsilon$ and $\epsilon < r < 20R$
mix and produce a gap.}
\label{latticethree}
\end{figure}

A cleaner picture of what is happening in the low energy region is
shown in Fig.\ref{latticethree}, where we have chosen to replot
both curves. This time we haven't rescaled the first $300$
eigenvalues and we have greatly expanded the vertical scale so as
to see what is happening near zero energy.  What one immediately
sees is that the branch of the plot which we identified as
belonging to the states localized on $r=\epsilon$ is in fact
massless, but the second branch, which we thought should look like
the spectrum of the $\lambda=0$ theory actually starts from a
small gap which lies just above the final value of the first
branch. How can this happen?  The answer seems to be that the term
which links the two distinct pieces of the Hamiltonian mixes what
would have been degenerate levels and splits them to produce a
continuously rising spectrum.  In particular this would suggest
that the lowest energy states we thought of as completely
localized at $r=\epsilon$ are in fact linear superpositions of
such states and low energy states of the $\lambda=0$ problem which
are not localized at all.  Clearly, it will require more work to
convert these results into a better understanding of how and where
the information which enters the horizon is really stored.

Having shown that analysis of the spectrum of the time dependent
Hamiltonian proves that the dynamics is time dependent, we wish to
conclude with a few remarks about the scenario this picture
suggests for the last moments of black hole evaporation. Clearly,
the picture strongly implied by our discussion is that during the
evaporation process a {\it remnant\/} is formed. This remnant
represents almost zero-energy degrees of freedom which can be
associated with the region $r=\epsilon$, since it is only by
including these degrees of freedom that we preserve the unitary
evolution of the system. The question therefore is what happens to
these degrees of freedom when the black hole evaporates.

One argument against a remnant would be that as holes
evaporate these remnants should announce their presence in
some dramatic fashion.  Since our treatment of the problem
encodes the changes in the metric into the  changes in the
coefficients appearing the field-theory Hamiltonian we can
approach this question by asking what really happens to
these coefficients during the final, rapid evaporation
phase.  If the coefficients of the Hamiltonian freeze into
the form they had shortly before the point at which the
methods used in this paper break down, then we see that
finally the system will be composed of two very weakly
coupled subsystems and the information stored in the
remnant will not burst out, but at best dribble out over
very long times. Obviously these remarks are not a proof
of anything. At best they describe an alternative scenario
for what might happen.  What we wish to emphasize is that
the Hamiltonian formulation of the problem coupled with
lattice techniques gives one a new way of probing these
issues.  Hopefully a better understanding of whether all
the information comes out will emerge as one studies this
problem in greater detail.

\section{Summary}

In the preceding sections of this paper we argued that a
consistent Hamiltonian formulation of the theory of a massless
scalar field in the background of a Schwarzschild black hole
exists, but at the expense of having an explicitly time-dependent
Hamiltonian.  We then reviewed the familiar discussion of the
moving mirror problem as a way of emphasizing that a perfectly
understood system with a time dependent Hamiltonian can evolve
unitarily and still some observers will think that they are in a
thermal bath of quanta.  The purpose of this review was to
emphasize that when one is dealing with a non-equilibrium system
and one imposes the ideas of equilibrium thermodynamics one can
come to surprising conclusions.

Following the discussion of the moving mirror problem, we returned
to the problem of the Schwarzschild black hole and used
essentially the same techniques to show that, with reasonable
assumptions about the initial state, our method leads to the usual
result of Hawking radiation.

There were two reasons for this discussion.  First, we wanted to
show that our treatment of the problem reproduces familiar
results. Second, we used it to argue that formulating the theory on a
spacelike slice at finite times clearly exhibits
the fact that the Hawking radiation phenomenon emerges under rather
general assumptions about the initial state on the quantization surface.
We included a discussion of the case of a massless scalar field
and an infalling reflecting mirror to sharpen this point.
Assuming that the mirror is static for large times in the past,  we
can reasonably argue  that it is proper to consider  the system to be
in the ground state of the theory, or in a state which differs
from it by a finite energy.  This naturally leads
us to the usual late time Hawking radiation.
Moreover, we saw that this Hawking radiation came from a place
on the original surface of quantization which is far from the mirror
and corresponds to the place from which null geodesics must depart
so as to be scattered from the mirror within an exponentially
small distance of the horizon.  Clearly this is not a special point
and its location depends upon when the mirror is allowed  to move.
Thus, we see that in order to get rid of the Hawking radiation
by suitably modifying the initial state would amount to making a
very strange ad-hoc assumption about the expectation values of
field operators on the initial quantization surface.

Having shown that our discussion leads to the usual picture of
what external observers would measure we turned to the question of
black hole entropy.  Once again  we discussed the case of a black
hole surrounded by a static mirror, but this time with the field
theory restricted to the inside of the sphere. We then argued that
from the point of view of an outside observer this system would
look like a classical body with a temperature equal to the Hawking
temperature and an equation of state which would imply, for an
equilibrium system, an entropy equal to the Hawking-Bekenstein
entropy.   Next, we looked inside the system and saw that it was
never in equilibrium and that it was in fact always in a pure
state.

Finally, we identified the ``information paradox'' with the question
of what is really happening at the spacelike singularity $r=0$.
By discretizing our time-dependent Hamiltonian and studying the
behavior of $H(\lambda)$ as a function of $\lambda$ we arrived
at a very interesting picture which implied both that the modes
at $r=0$ and the modes going out to large Schwarzschild $r$
mix.  This left us with the possibility that the endpoint of
the evolution of the hole when it evaporates is a field theoretic
system in which some of the information is stored in a very weakly
coupled (but not decoupled) remnant and some has been squeezed out as it
approached $r=0$.  Clearly the questions raised in this section of
the paper by far outnumber the results and these issues merit further
study.

In conclusion, we would like to reiterate the point made in the introduction
to the paper.  Our studies show that the theory of a massless scalar field
in a black hole background is perfectly consistent with unitary
time evolution and that somehow the theory resolves any
supposed paradoxes in its own way.  This led us to the
conclusion that this semi-classical problem alone does not provide
any smoking gun telling us what a correct theory of quantum gravity
must look like.

{\em Acknowledgements:}
We would like to thank James Bjorken for useful conversations.
This research was supported by the DOE under grant number
DE-AC03-76SF00515.

\end{document}